\documentclass[twocolumn]{aastex62}
\usepackage{url}
\usepackage{gensymb}
\usepackage{textcomp}
\usepackage{xspace}
\usepackage{natbib,amsmath} \usepackage{graphicx}
\usepackage{hyperref} \usepackage{float} \usepackage{subfigure}
\hypersetup{colorlinks=true}
\usepackage{tabularx}
\providecommand{\e}[1]{\ensuremath{\times 10^{#1}}}

\usepackage{array}

\begin{document}
\title{The long-term evolution and appearance of Type Iax postgenitor stars}

\author[0000-0002-0659-1783]{Michael Zhang}
\affil{Department of Astronomy, 
	California Institute of Technology, Pasadena, CA 91125, USA}
\author[0000-0002-4544-0750]{Jim Fuller}
\affil{Department of Astronomy, 
	California Institute of Technology, Pasadena, CA 91125, USA}
\author[0000-0002-4870-8855]{Josiah Schwab}
\altaffiliation{Hubble Fellow}
\affil{Department of Astronomy and Astrophysics, 
	University of California Santa Cruz, Santa Cruz, CA 95064, USA}
\author{Ryan J. Foley}
\affil{Department of Astronomy and Astrophysics, 
	University of California Santa Cruz, Santa Cruz, CA 95064, USA}
\correspondingauthor{Michael Zhang}	
\email{mzzhang2014@gmail.com}

\begin{abstract}
Type Iax supernovae may arise from failed explosions of white dwarfs that leave behind a bound remnant (i.e., a ``postgenitor" star) that could be identified in wide field surveys. To understand their observational signatures, we simulate these white dwarf (WD) postgenitors from shortly after explosion until they move back down the WD cooling track, and we consider several possible WD masses and explosion energies. To predict the peculiar surface abundances of the WD postgenitors, our models take into account gravitational settling and radiative levitation. We find that radiative levitation is significant at temperatures above a mass-dependent critical temperature, typically in the range $T_{\rm eff} \approx 50-100 \times 10^3\,{\rm K}$, significantly increasing surface abundances of iron-group elements.  Due to enhanced iron group opacity compared to normal WDs, the postgenitor peak luminosity and cooling timescale depend sensitively on mass, with more massive WDs becoming brighter but cooling much faster.  We discuss our results in light of recently discovered hypervelocity white dwarfs with peculiar surface compositions, finding that our low-mass postgenitor models match many of their observational characteristics. Finally, we explore the effects of thermohaline diffusion, tentatively finding that it strongly suppresses abundance enhancements created by radiative levitation, but more realistic modeling is required to reach a firm conclusion.
\end{abstract}



\section{Introduction}
\label{sec:introduction}
In 2002, a very peculiar supernova (SN) was discovered at Palomar Observatory \citep{wood-vasey_2002}.  This supernova, soon to be named 2002cx, did not fit neatly into any category.  It had the pre-maximum spectrum similar to that of a Ia, with absorption lines from intermediate-mass elements and iron.  However, it also had a very low luminosity, half the typical expansion velocity, and atypically red colors.  The discovery paper \citep{li_2003}, aptly entitled `SN 2002cx: The Most Peculiar Known Type Ia Supernova', concluded that no existing model can explain the supernova.

Since then, dozens of supernovae like 2002cx have been discovered.  \cite{foley_2013} grouped these events into a distinct class, called Type Iax, and estimate its event rate as roughly 1/3 that of type Ia. SNe Iax are the most common class of peculiar Ia-like supernovae \citep{jha_2017}.  Like their prototype, type Iax supernovae are characterized by a Ia-like pre-maximum spectrum, but with peak luminosities typically a few magnitudes fainter and expansion velocities a few times slower.  SNe Iax are more inherently diverse than SNe Ia, with no strong width-luminosity relation, peak luminosity varying over 4-5 magnitudes, and velocities varying by a factor of 4. SNe Iax also have other interesting properties.  For example, they have never been observed in elliptical galaxies (although there is one example in an S0 galaxy: \citealt{foley_2010}) and prefer star-forming spirals, yet there is also no sign of ongoing star formation at the site of any Iax \citep{foley_2013}.

Ia/Iax SNe result from the explosion of a white dwarf with a binary companion, but neither the nature of the companion nor the mechanism that triggers the explosion is known.  In the single degenerate Ia/Iax scenario, the companion is a non-degenerate star; in the double degenerate Ia/Iax scenario, it is another white dwarf.  \cite{wang_2013} argue for a single degenerate scenario for SNe Iax, arguing that they could be the product of accretion from a helium star onto a CO WD, a mechanism which reproduces the long delay times and luminosity diversity of SNe Iax.  The detection of a luminous blue progenitor of a SN Iax \citep{mccully_2014} supports this scenario.  It has also been proposed that SNe Iax could result from hybrid CONe progenitors \citep{meng_2014}. \cite{bravo_2016} investigates explosions of hybrid CONe WDs created by off-center carbon burning in intermediate mass stars, finding such explosions could leave behind bound remnant WDs.  \cite{kashyap_2018} explain Ia/Iax SNe as being due to the merger of a CO WD with a ONe WD.  While the CO mixture burns easily, the ONe mixture does not, creating a low-luminosity transient with small ejecta mass.

Despite these unknowns, the leading explanation for SNe Iax is that they are white dwarf (WD) deflagrations that do not lead to detonations \citep{branch_2004, jha_2006, phillips_2007}, which explains the Ia-like spectrum, the low luminosity, and the low expansion velocity.  Deflagrations also tend to produce a wide range of explosion energies, explaining the diversity of type Iax events, whereas Chandrasekhar-mass detonations are more uniform in their properties.  \cite{kromer_2013} performed a 3D deflagration simulation of a Iax explosion, successfully reproducing characteristic observational features of SN 2005hk in the optical and near-infrared. The asymmetric mass ejection may also impart a kick of several hundred km/s to the bound remnant star \citep{jordan_2012}.

If SNe Iax supernovae are truly partial deflagrations of CO WDs or deflagrations and delayed detonations of hybrid CONe cores, they will not be energetic enough to unbind the white dwarf.  \cite{shen_schwab_2017} simulate bound WD postgenitors using the MESA stellar evolution code \citep{paxton_2011}.  They take into account delayed $^{56}$Ni decay and super-Eddington winds to predict the light curve from days after the explosion to 1000 years afterwards.  They report high uncertainties, but reasonably good matches for the late-time luminosity, temperature, and velocity of SN 2005hk, SN 2008A, and SN 2008ha.

Recently, \cite{vennes_2017} discovered a hypervelocity white dwarf (LP 40-365) with an oxygen-neon atmosphere and abundant intermediate-mass elements; they consider LP 40-365 a candidate for a Iax postgenitor.  Using Gaia data, \cite{raddi_2018} confirm the hypervelocity nature of the object while reporting a radius of $0.18 \pm 0.01 R_{\odot}$ and a mass of $0.37_{-0.17}^{+0.29} M_{\odot}$. \cite{shen_2018} discovered a few more peculiar hypervelocity subdwarf/white dwarf stars with similar properties. These objects also have no detectable hydrogen or helium, but do have strong carbon, oxygen, iron, magnesium, and calcium features.  The large space velocities, peculiar surface compositions, and unusual masses/radii suggest these stars could be Iax postgenitors, although those authors posit that they are the degenerate donor star companions of white dwarfs that exploded as SN Ia.  Finally, \cite{kepler_2016} reports an enigmatic WD with a nearly pure oxygen atmosphere.  It is unclear how this object formed, but we speculate that it could be a ONe or CONe WD that deflagrated long ago.

In this paper, we extend the postgenitor simulations of \cite{shen_schwab_2017} to very late times to determine the properties of these bound remnant stars.  We discuss the setup of our simulations in Section~\ref{sec:models}, show the most salient characteristics in Section~\ref{sec:results}, and analyze the results while comparing them to hypervelocity stars like LP 40-365 in Section~\ref{sec:discussion}.

\section{Models}
\label{sec:models}

To simulate WD postgenitors, we perform stellar evolution calculations using MESA \citep{paxton_2011, paxton_2013, paxton_2015, paxton_2018} version 10000. We set up a grid of models with different initial conditions, then evolve them hydrostatically to predict the evolution of observables such as temperature, luminosity, radius, and surface abundances.  

We do not attempt to simulate the supernova itself, nor do we take into account binary interaction, winds, radioactive decay, or detailed radiative transfer.  Our simulation is meant to start at late times (more than $\sim$100 yr after explosion), when all significant radioactive nuclides have decayed, all winds have died down, and the supernova remnant has long since become optically thin.  The simulation assumes hydrostatic quasi-equilibrium at all times, taking into account convection, radiative transport, element diffusion, radiative acceleration, and neutrino cooling to evolve the white dwarf until it is far down its cooling track.

\subsection{Initial conditions}

To set up somewhat realistic initial conditions, we take a packaged white dwarf model from MESA and adjust its properties.  We relax its composition to one appropriate for carbon-oxygen Iax postgenitors \citep{kromer_2013}, then relax the outer portions of the white dwarf to a constant entropy.  These outer portions, which will henceforth be called the ``envelope'', represent a combination of the nuclear ashes and the fall-back ejecta.  A homogenous and constant entropy envelope is expected in regions well mixed by convective burning, which enforces a nearly isentropic structure, and is likely to be the case in the thermally supported envelope of the WD soon after deflagation.

To account for the pollution with nucleosynthetic burning products of the deflagation, we relax the elemental abundances in the stellar envelope to those shown in Table \ref{table:composition}.  These numbers are taken from \cite{kromer_2013}, which describes 3D deflagration simulations of Iax supernovae and predicts final elemental abundances of both the postgenitor and the ejecta.  The precise numbers are not very important for our purposes, as we are more interested in the evolution of surface abundances over time due to diffusion and radiative levitation. However, we note that the large iron-group abundance, roughly 30x solar, is important for the evolution of the WD and its spectroscopic appearance.

To account both for the inherent diversity of Iax supernovae and for uncertainty in the explosion process and its outcome, we use a grid of 24 initial conditions.  The grid contains 4 postgenitor masses (0.15, 0.3, 0.6, 1 $M_{\odot}$), 3 envelope fractions (10, 50, 90\% by mass), and 2 envelope specific entropies (3\e{8}, 5\e{8} erg g\textsuperscript{-1} K\textsuperscript{-1}).  The postgenitor masses and envelope fractions were chosen to encompass a large range of possibilities for the explosion process.  If the explosion is violent and ejects a large amount of mass, for example, one might expect a small postgenitor mass.  If significant burning takes place but does not result in large amounts of ejecta, one might expect a large envelope fraction.  The final parameter, the specific envelope entropy, was chosen so that the higher entropy corresponds to unbound or very loosely bound white dwarfs, while the lower entropy always corresponds to bound objects.

\begin{table*}[t]
  \centering
  \caption{Composition of SN Iax postgenitor envelopes}
  \begin{tabular}{C c}
  \hline
  	{\rm Isotope} & Mass abundance (\%)\\
      \hline
      ^{12}{\rm C}  &	42\\
      ^{16}{\rm O}	&	48\\
      ^{20}{\rm Ne}	&	5.3\\
      ^{24}{\rm Mg}	&	0.4\\
      ^{28}{\rm Si}	&	1.5\\
      ^{32}{\rm S}	&	0.4\\
      ^{40}{\rm Ca}	&	0.03\\
      ^{56}{\rm Fe}	&	3.6\\
      ^{58}{\rm Ni}	&	0.3\\      
      \hline
  \end{tabular}
  \label{table:composition}
\end{table*}

\subsection{Input physics}
After white dwarfs are set up with the appropriate initial conditions, we evolve them forward in time with MESA.  Our inlist is provided in Appendix A, but here we describe important settings.  We use Type 2 opacities derived from the Opacity Project, OP \citep{seaton_1995, seaton_2005}, and enable both diffusion and radiative levitation of all elements being simulated.  Furthermore, we restrict the network of isotopes and reactions to include only those isotopes we simulate and no reactions, because the isotopes we added are not expected to undergo nuclear reactions.  This prevents numerical errors from creating spurious elements, which levitation or diffusion might then concentrate--a phenomenon we had previously seen in our models.

In addition to diffusion, radiative acceleration, and convection, we introduce an additional source of mixing with \texttt{min\_D\_mix=1.0}.  This minimum diffusion coefficient of 1 $\rm cm^2\,s^{-1}$ ameliorates numerical problems, such as unphysically sharp composition gradients and unrealistically rapid composition fluctuations in the outer layers of the envelope.  In physical terms it may correspond to sources of mixing not accounted for in our model, such as rotational mixing.  \texttt{min\_D\_mix} is always irrelevant in convective zones, but its existence prevents the photosphere from becoming pure carbon at very late times, whereas a pure carbon atmosphere is theoretically expected at very late times.

One important effect which we neglect is thermohaline diffusion caused by inverse composition gradients that can result from radiative levitation.  Thermohaline diffusion will counteract the effects of radiative levitation and may be enough to flatten the composition gradient almost completely.  However, thermohaline diffusion is difficult to properly model in combination with strong radiative acceleration, especially near the stellar photosphere.  We discuss the difficulties in Subsection \ref{subsec:thermohaline}, and describe the outcome of our exploratory thermohaline simulation.

Radiative acceleration can be calculated using the OP module in MESA, which uses element-specific opacity data from the Opacity Project.  We found the existing module to suffer multiple computational problems, including bugs and slow run speeds.  Interpolation problems near the edges of the OP opacity grid also caused catastrophic failures.  As a consequence, we have rewritten large portions of the module to improve performance and resilience.  These steps include deleting most of the code, which calculates parameters that are then thrown away; changing the interpolation algorithm; refactoring code to reduce duplication; and implementing a cache.  We are in the process of contributing some of these improvements back into MESA, but in the meanwhile, our custom version of MESA can be downloaded at \url{http://www.astro.caltech.edu/~mz/custom_mesa_10000.tar.gz}.

The algorithm for calculating radiative accelerations is given in \cite{seaton_2005}.  To summarize, the radiative acceleration for element $k$ is:
\begin{align}
g_{{\rm rad}, k} &= \frac{F}{c}\frac{\mu}{\mu_k}\kappa_{R}\gamma_k\\
\gamma_k &= \int{\frac{\sigma_k^{mta}(u)}{\sigma(u)} \frac{F(u)}{1-e^{-u}}du}\\
\sigma_k^{mta} &= \sigma_k(u)(1-e^{-u}) - a_k(u)
\end{align}

Here:

\begin{itemize}
    \item $u = \frac{h\nu}{k_BT}$,
    \item $\sigma_k(u)$ is the monochromatic cross section for element $k$ at the wavelength corresponding to $u$,
    \item $\sigma(u)$ is the cross section for the mixture,
    \item $\sigma_k^{mta}(u)$ is the cross section for momentum transfer to the atom,
    \item $a_k(u)$ is a momentum transfer correction factor,
    \item $k_R$ is the Rosseland mean opacity,
    \item $\mu$ is the mean atomic weight of the mixture,
    \item $F(u)$ is the flux (assuming a blackbody),
    \item $\gamma_k$ is an integrated cross section ratio.
\end{itemize}

The cross sections and correction factors are both provided by the Opacity Project for 10,000 wavelengths.  For each element and each wavelength, OP provides these quantities in a non-rectangular grid whose dimensions are temperature and electron density ($n_e$).  

However, electron density is not a useful variable because MESA uses temperature and density to define the conditions in a zone, not temperature and electron density.  Electron density can be converted to physical density through another grid provided by OP, namely that of electrons per atom for each element at each T and $n_e$:
\begin{align}
\rho &= \frac{n_e}{e_{\rm avg}} \mu\\
&= \frac{n_e \mu}{\sum_k f_k e_k(T, n_e)}
\end{align}
where $e_{\rm avg}$ is the number of free electrons per atom for the mixture, in which each element $k$ has abundance $f_k$.

To calculate the radiative acceleration of a zone, we first calculate the physical density of every point on the T-$n_e$ grid, given the composition of the zone.  We then find the 16 grid points closest to the zone's temperature and density, and calculate the radiative acceleration for those 16 points.  Cubic interpolation is then used to calculate the acceleration at the zone's temperature and density:
\begin{align}
g_{\rm rad}(T, \rho) = \sum_{i=1}^4 \sum_{j=1}^4 (\log T)^i (\log \rho)^j c_{ij},
\end{align}
where the 16 $c_{ij}$ constant coefficients are derived by fitting to the 16 grid points.  We use cubic interpolation instead of linear interpolation in order to preserve the continuity of the derivatives of opacity with respect to temperature and pressure, which MESA requires for its hydrostatic solver.

\section{Results}
\label{sec:results}

\subsection{Overview}
Our grid of 24 models resulted in a diversity of outcomes, listed in Table \ref{table:outcomes}.  In the majority of cases, the simulation followed a canonical pattern, explained in more detail in Section \ref{subsec:canonical}.  The WD would initially cool and dim.  After some time, there is a rapid re-brightening and reheating event.  Depending on the peak temperature, radiative levitation could become important at this stage, creating an atmosphere dominated by nickel and iron.  Afterwards, the WD cools, radiative levitation fails, and gravitational settling takes over.  The WD enters onto the cooling track and follows it thereafter.

\begin{table*}[t]
  \centering
  \caption{Simulation outcomes}
  \begin{tabular}{c c c c}
  \hline
  	Mass ($M_{\odot}$) & Envelope entropy (erg g\textsuperscript{-1} K\textsuperscript{-1}) & Envelope fraction & Outcome\\
      \hline
      0.15 & 3\e{8} & 0.1 & Normal\\
      0.15 & 3\e{8} & 0.5 & Normal\\
      0.15 & 3\e{8} & 0.9 & Normal\\
      0.15 & 5\e{8} & 0.1 & Normal\\
      0.15 & 5\e{8} & 0.5 & Unbound, but evolved to end\\
      0.15 & 5\e{8} & 0.9 & Unbound\\
      0.3 & 3\e{8} & 0.1 & Normal\\      
      0.3 & 3\e{8} & 0.5 & Normal\\
      0.3 & 3\e{8} & 0.9 & Normal\\
      0.3 & 5\e{8} & 0.1 & Normal\\
      0.3 & 5\e{8} & 0.5 & Super-Eddington\\
      0.3 & 5\e{8} & 0.9 & Unbound\\
      0.6 & 3\e{8} & 0.1 & Normal\\
      0.6 & 3\e{8} & 0.5 & Normal\\
      0.6 & 3\e{8} & 0.9 & Normal\\
      0.6 & 5\e{8} & 0.1 & Super-Eddington\\
      0.6 & 5\e{8} & 0.5 & Unbound\\
      0.6 & 5\e{8} & 0.9 & Normal\\
      1 & 3\e{8} & 0.1 & Normal\\
      1 & 3\e{8} & 0.5 & Normal\\
      1 & 3\e{8} & 0.9 & Normal\\
      1 & 5\e{8} & 0.1 & Super-Eddington\\
      1 & 5\e{8} & 0.5 & Super-Eddington\\
      1 & 5\e{8} & 0.9 & Normal\\
      \hline
  \end{tabular}
  \label{table:outcomes}
\end{table*}

\subsection{Canonical case}
\label{subsec:canonical}

\begin{figure}
  \includegraphics [width= 0.5\textwidth]{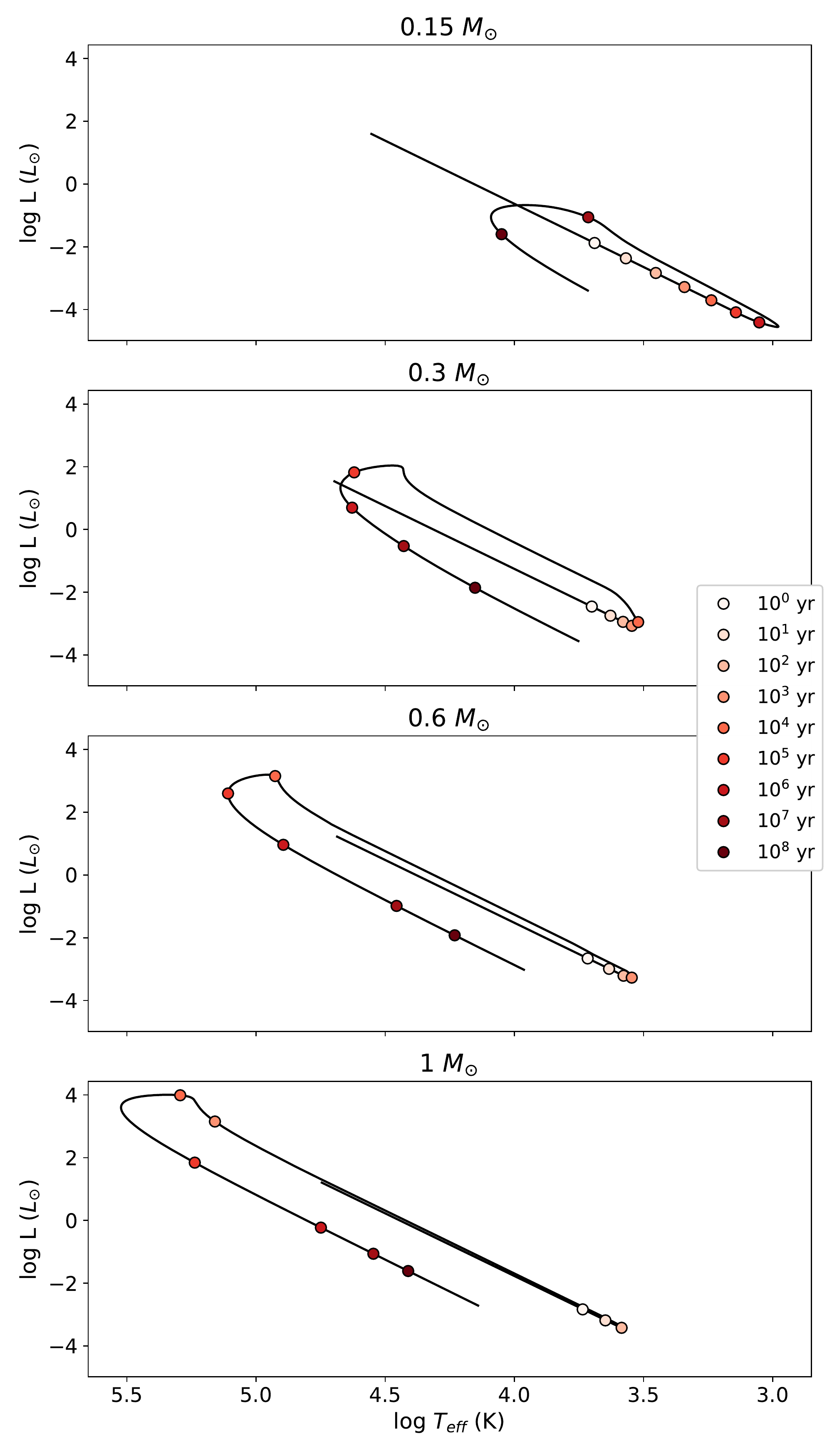}
  \caption{HR diagrams for different masses at the same envelope fraction (50\%) and envelope entropy (3\e{8} erg g\textsuperscript{-1} K\textsuperscript{-1}).}
\label{fig:HR_diagrams}
\end{figure}

Figure \ref{fig:HR_diagrams} illustrates the behavior of a typical postgenitor in our simulations.  The initial cooling and dimming is due to the outer layers, which have a short thermal timescale, radiating away heat.  During this phase the outer envelope is convective--the constant entropy envelope exactly fulfills the Schwartzchild criterion, and preferential cooling of the outer layers only increases the temperature gradient.  Although there is abundant heat buried deeper in the envelope, this heat has not yet had time to diffuse out.  When the heat does diffuse out, it results in the reheating and rebrightening event seen in all three diagrams.  The WD becomes very hot and bright.  At some point it begins to cool again, following the normal cooling track for WDs.

A few features of Figure \ref{fig:HR_diagrams} are worth noting.  First, the final cooling track is to the left of the initial cooling track, and also to the left of the rebrightening track.  Since $L = 4\pi R^2 \sigma T_{\rm eff}^4$ by definition, this leftward shift indicates a substantial decrease in radius.  Second, higher masses lead to higher temperatures and higher luminosity.  

\begin{figure}
  \includegraphics [width= 0.5\textwidth]{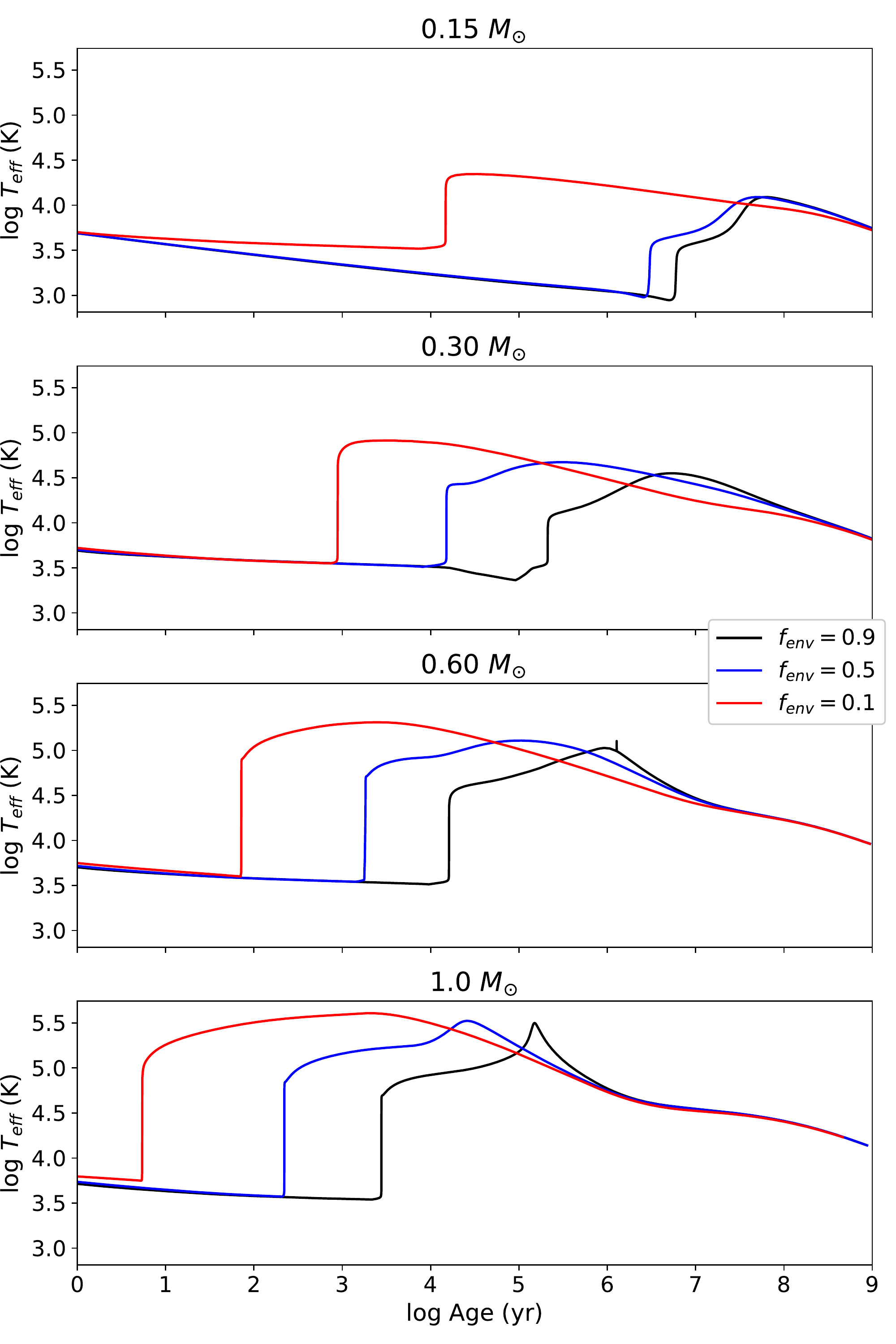}
  \caption{Effective temperature as a function of age for all low entropy scenarios.}
\label{fig:log_Teff_vs_time}
\end{figure}

\begin{figure}
  \includegraphics [width= 0.5\textwidth]{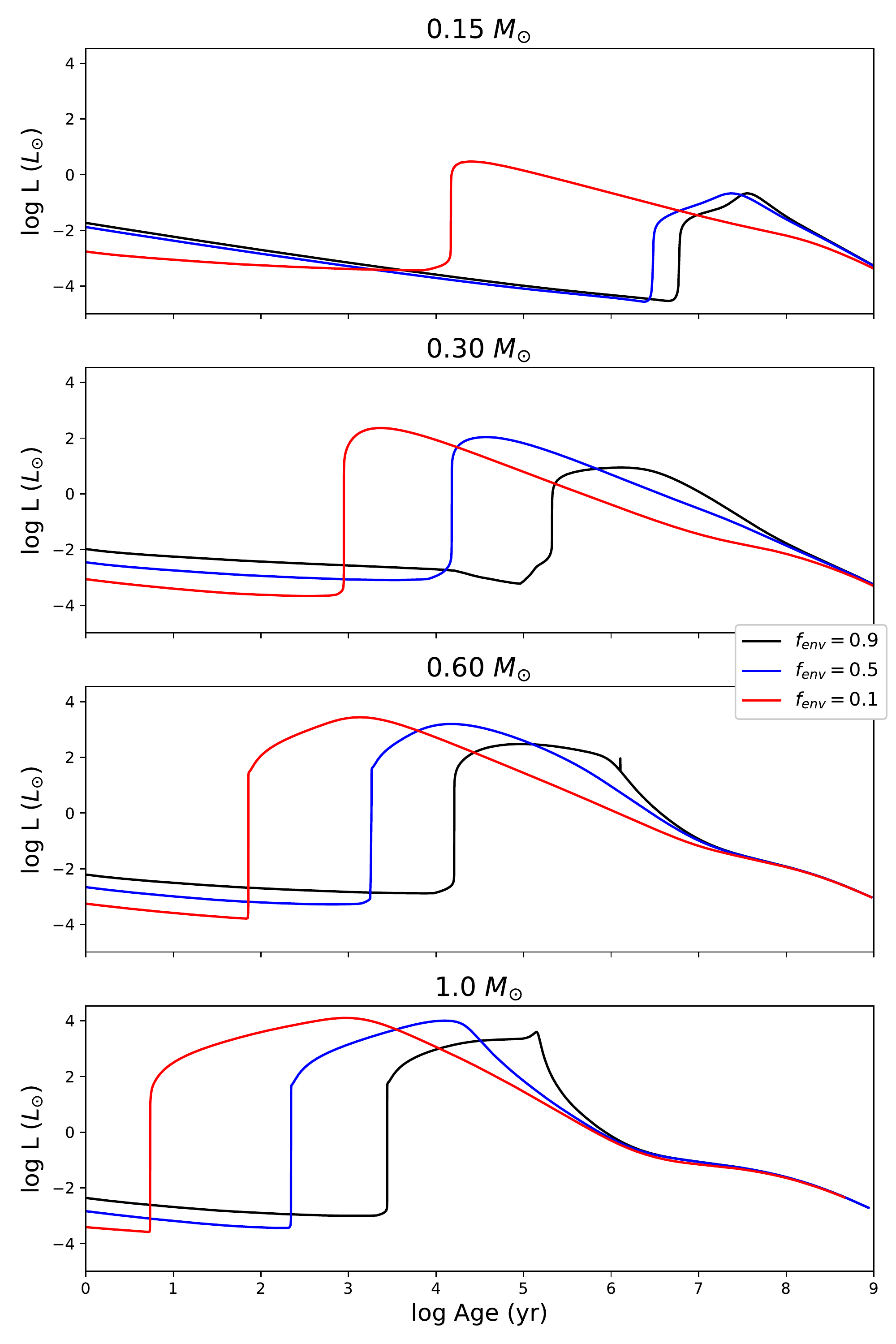}
  \caption{Luminosity as a function of age for all low entropy scenarios.}
\label{fig:log_L_vs_time}
\end{figure}

Figure \ref{fig:log_Teff_vs_time} and Figure \ref{fig:log_L_vs_time} show the time evolution of temperature and luminosity for all low entropy models.  As can be seen, all scenarios follow the canonical pattern of dimming, rapid re-brightening, and re-dimming.  The timescales, however, are drastically different.  Higher-mass WDs evolve much faster, as do WDs with low envelope fraction.

\begin{figure}
  \includegraphics [width= 0.5\textwidth]{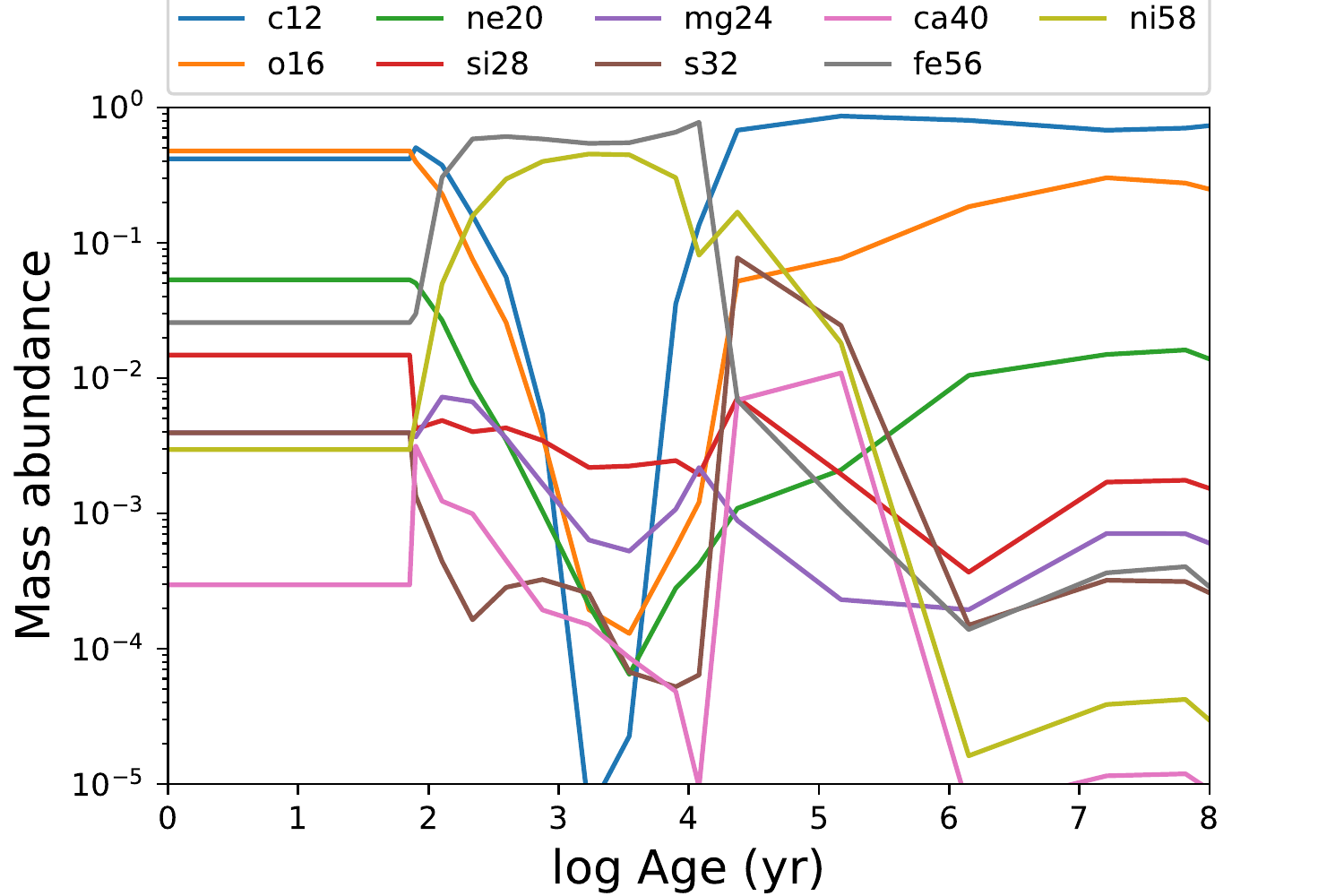}
  \caption{Surface abundances as a function of age.  Plotting 0.6 $M_{\odot}$, 10\% envelope fraction, low entropy scenario.  At very early times ($<$ 100 yr), the white dwarf composition does not change because there is a convection zone extending to the surface.  Radiative levitation is significant from 100 to 10,000 yr, after which the white dwarf has cooled enough for gravitational settling to take over.}
\label{fig:surface_abundances}
\end{figure}

Figure \ref{fig:surface_abundances} shows the surface abundances of all elements over time in one specific model, namely the one with WD mass 0.6 $M_{\odot}$, 10\% envelope fraction and lower envelope entropy.  During the initial cooling stage, surface abundances are constant due to the convective zone in the envelope.  In fact, the convective zone ensures that the entire WD has near-uniform composition throughout this stage. After the re-brightening event at $\sim$70 yr, the envelope becomes radiative, and the the high surface temperatures cause radiative acceleration of iron and nickel toward the surface.  These two elements are preferentially levitated because they have a large number of lines, thus fulfilling $g_{\rm rad} > g$.  As the postgenitor cools, radiative levitation eventually fails to hold heavy elements aloft, and they fall out of the photosphere.  Gravitational settling then takes over, interacting with the artificially injected mixing (min\_D\_mix = 1) to create stable surface abundances after 1 Myr.

\subsection{Abnormal cases}

\begin{figure}
  \centering \subfigure[HR evolution] {\includegraphics
    [width=0.5\textwidth]{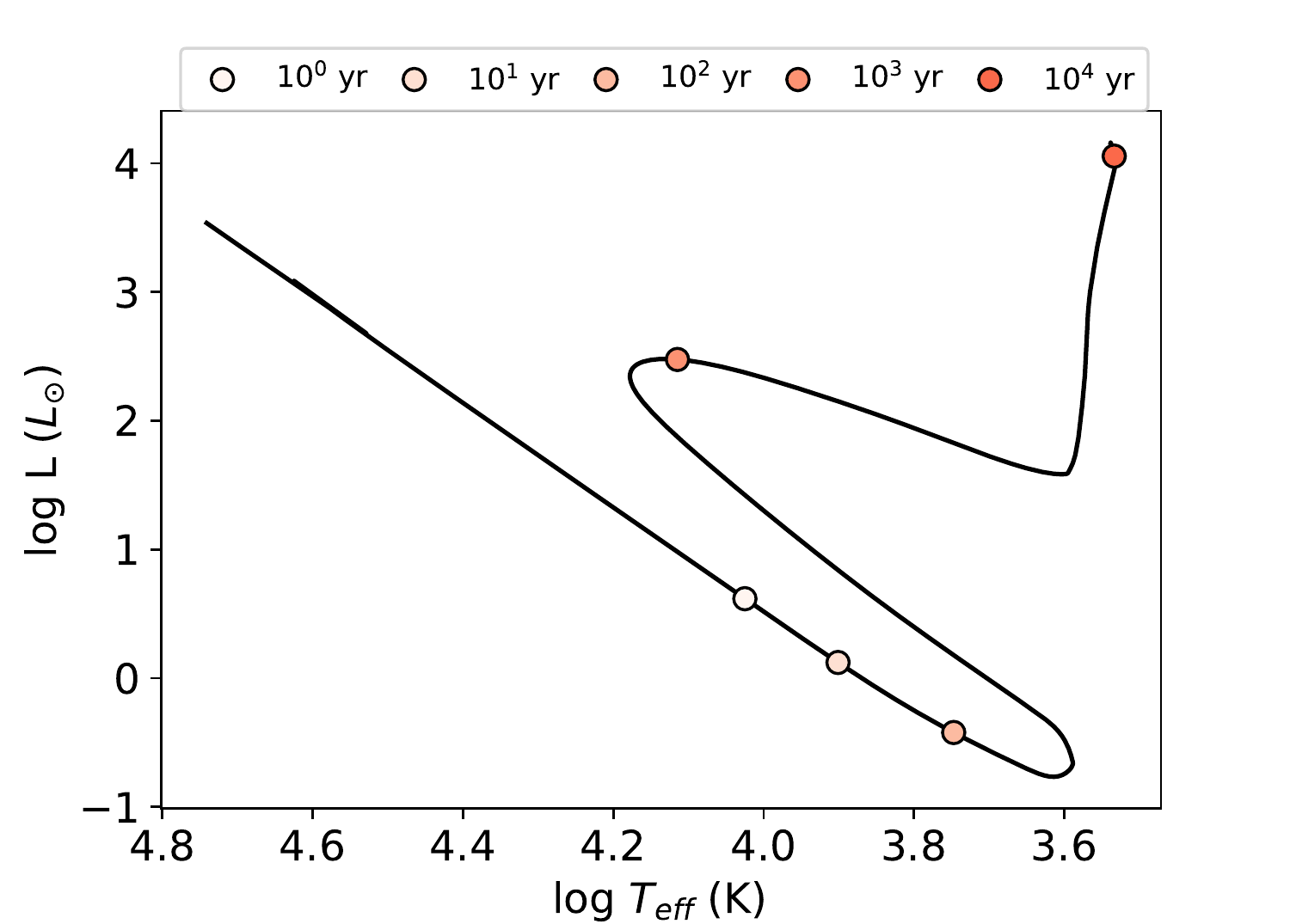}}\qquad
  \subfigure[Radius vs age] {\includegraphics
    [width=0.5\textwidth]{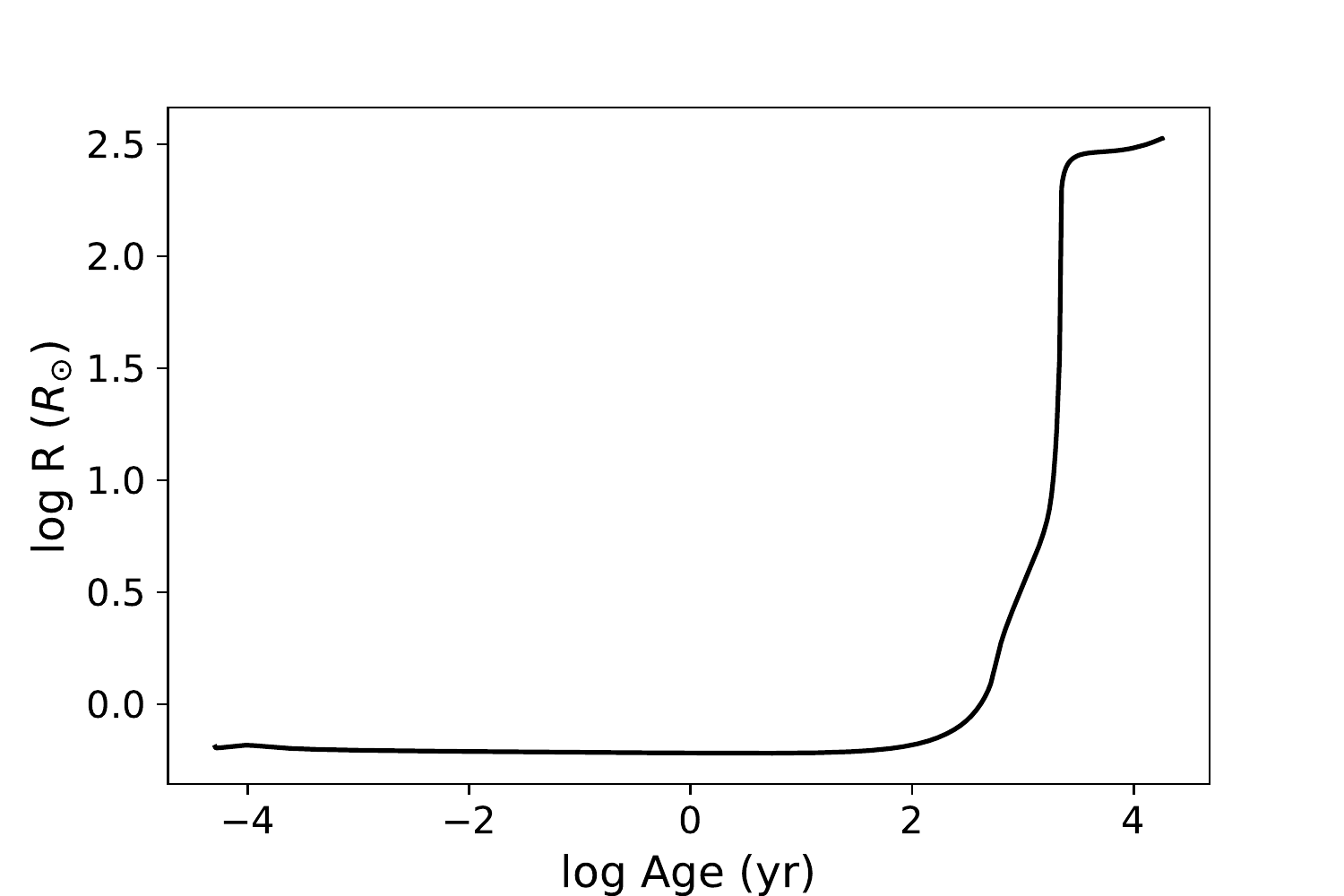}}
    \caption{A particularly interesting atypical result.  This is the 1 $M_{\odot}$, 50\% envelope fraction, high envelope entropy model.}
    \label{fig:atypical_case}
\end{figure}

The abnormal cases are simulations that fail to reach the cooling track.  In some cases this is because the model has positive total energy, and is therefore not gravitationally bound.  This occurs in the low mass WDs with high entropies and high envelope fractions.  In other cases the model is gravitationally bound, but has enough energy in the envelope that the luminosity exceeds Eddington luminosity, the envelope expands to tens or hundreds of solar radii, and MESA stalls.  This occurs preferentially in the high mass WDs with thin envelopes, as high mass WDs have enough gravity to keep the envelope bound.  In both cases significant mass loss is expected, though we do not attempt to calculate such mass loss here.  One particularly interesting example of the second case is shown in Figure \ref{fig:atypical_case}, where the red dwarf dims, rebrightens, dims again, and rebrightens again for the final time while expanding into a red giant.  Not surprisingly, all of these abnormal simulations occur when the envelope entropy is high.

These pathological cases are interesting in their own right. There is no reason, for example, why a surviving WD cannot have an ultra-hot envelope, or why the envelope cannot puff up and drive a wind.  Indeed, \cite{foley_2016} find P-Cygni features on permitted lines at late times (t $>$ 200 d) in a Type Ia SN, implying an expanding envelope that the authors attribute to a super-Eddington wind.  The implied velocity of 410 km \textsuperscript{-1} is consistent with the escape velocity of a $R_{\odot}$ postgenitor.  Furthermore, the narrow forbidden lines have a similar velocity as the expanding photosphere, implying they are also due to the wind.

Unfortunately, it is difficult to simulate these highly inflated mass-losing objects, because MESA runs into numerical difficulties when mass becomes unbound or when the luminosity becomes super-Eddington near the photosphere. Although it is possible to introduce a wind, there is no guarantee that existing wind prescriptions--developed for RGB and AGB stars--will be suitable for these peculiar objects. There is also no guarantee that a wind would help with the convergence problems.  \cite{lau_2012} model a similar instability associated with super-Eddington winds in AGB stars, but their simulation also crashed due to numerical problems.  Thus, while these hyper-inflated carbon/oxygen are interesting, our MESA models are not trustworthy representations, so we set aside these pathological cases and focus on the `Normal' outcomes in Table \ref{table:outcomes}.

\section{Discussion}
\label{sec:discussion}

In the previous section, we presented the salient characteristics of our simulations.  It is worth discussing which aspects of our simulations are believable and which should be taken with a grain of salt.  After this discussion, we will compare our simulation results to the observations of LP 40-365, a candidate Iax postgenitor.

\subsection{Decline-rise-decline pattern}

\begin{figure*}
  \centering \subfigure {\includegraphics
    [width=0.5\textwidth]{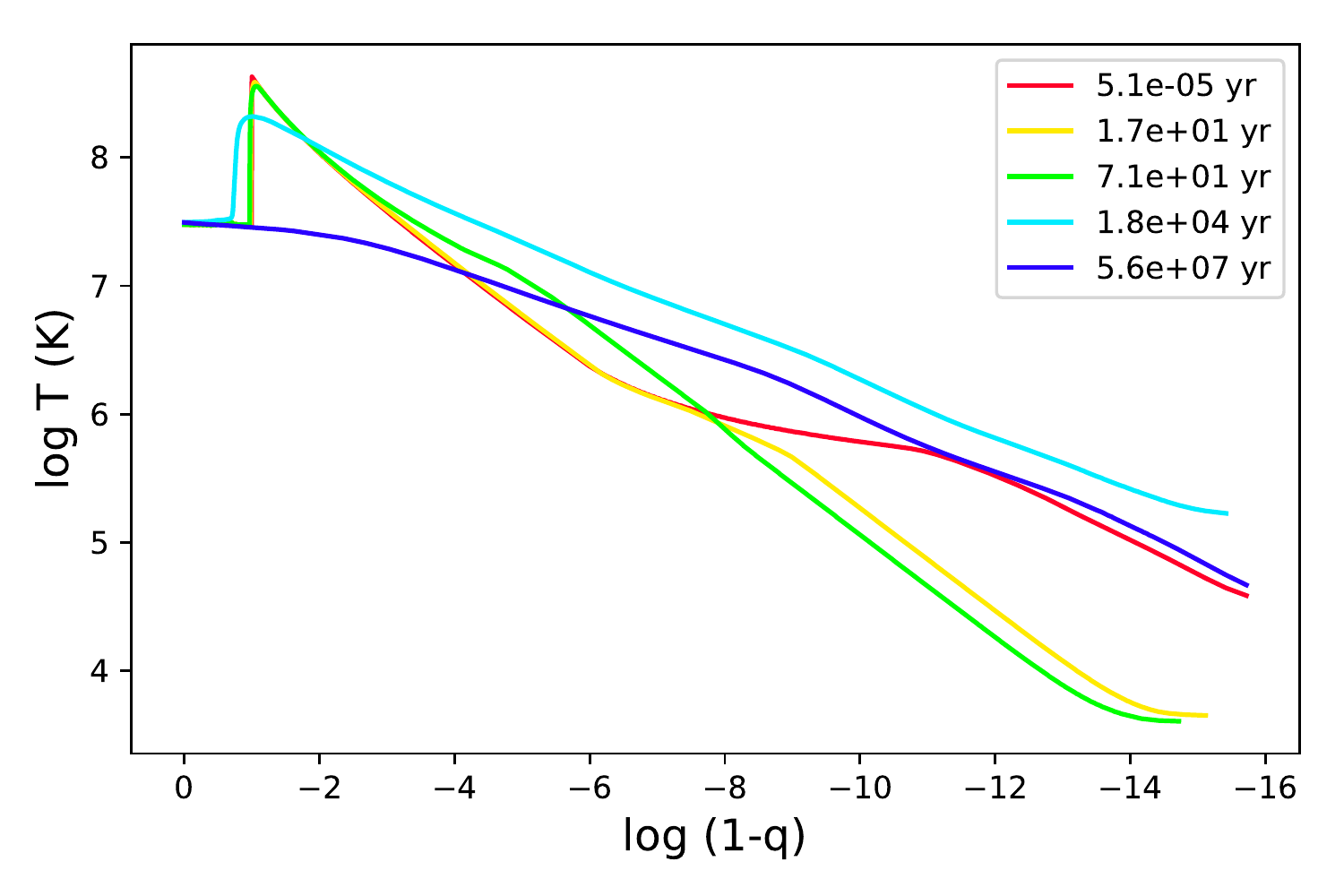}}\subfigure {\includegraphics
    [width=0.5\textwidth]{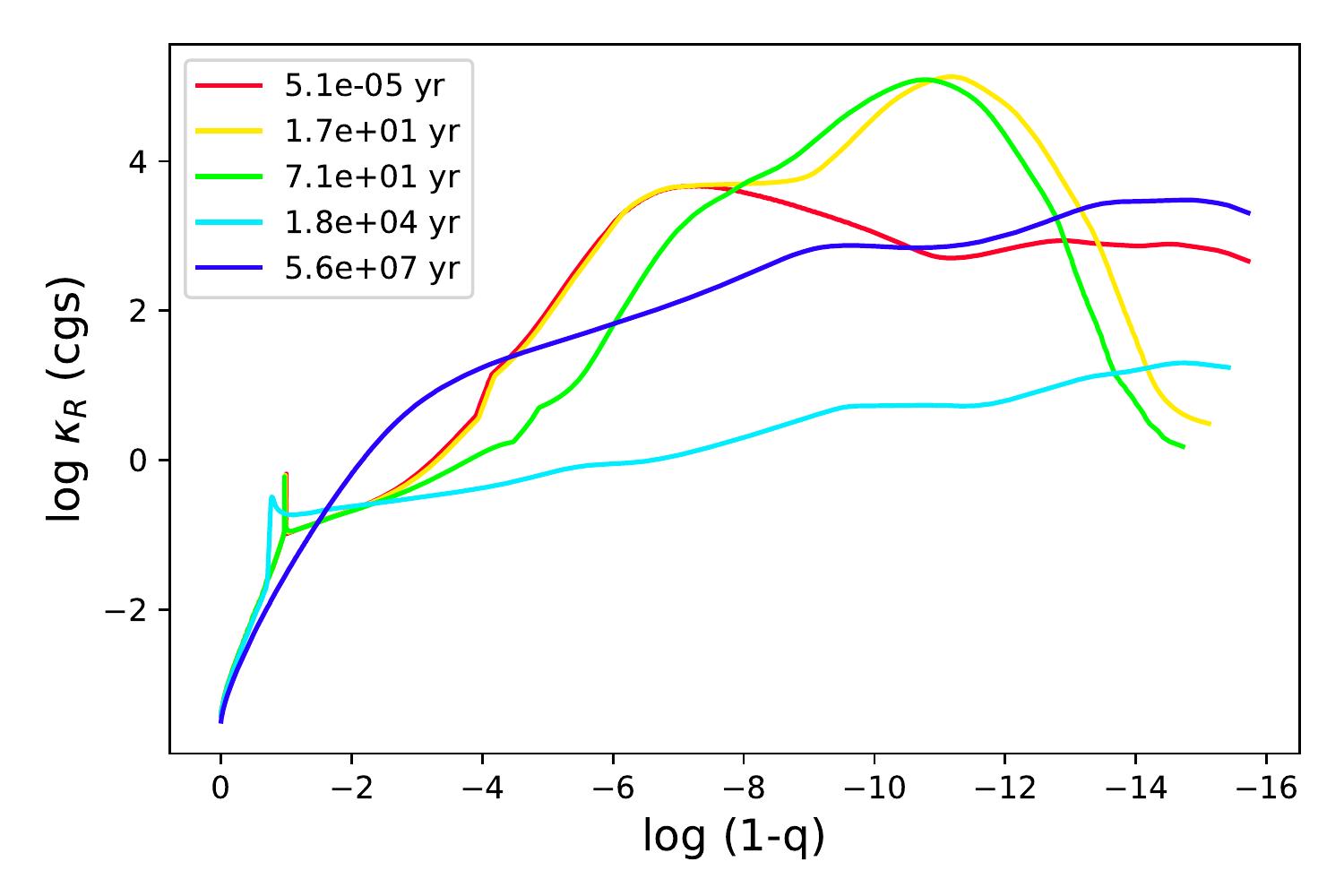}}
\subfigure{\includegraphics
    [width=0.5\textwidth]{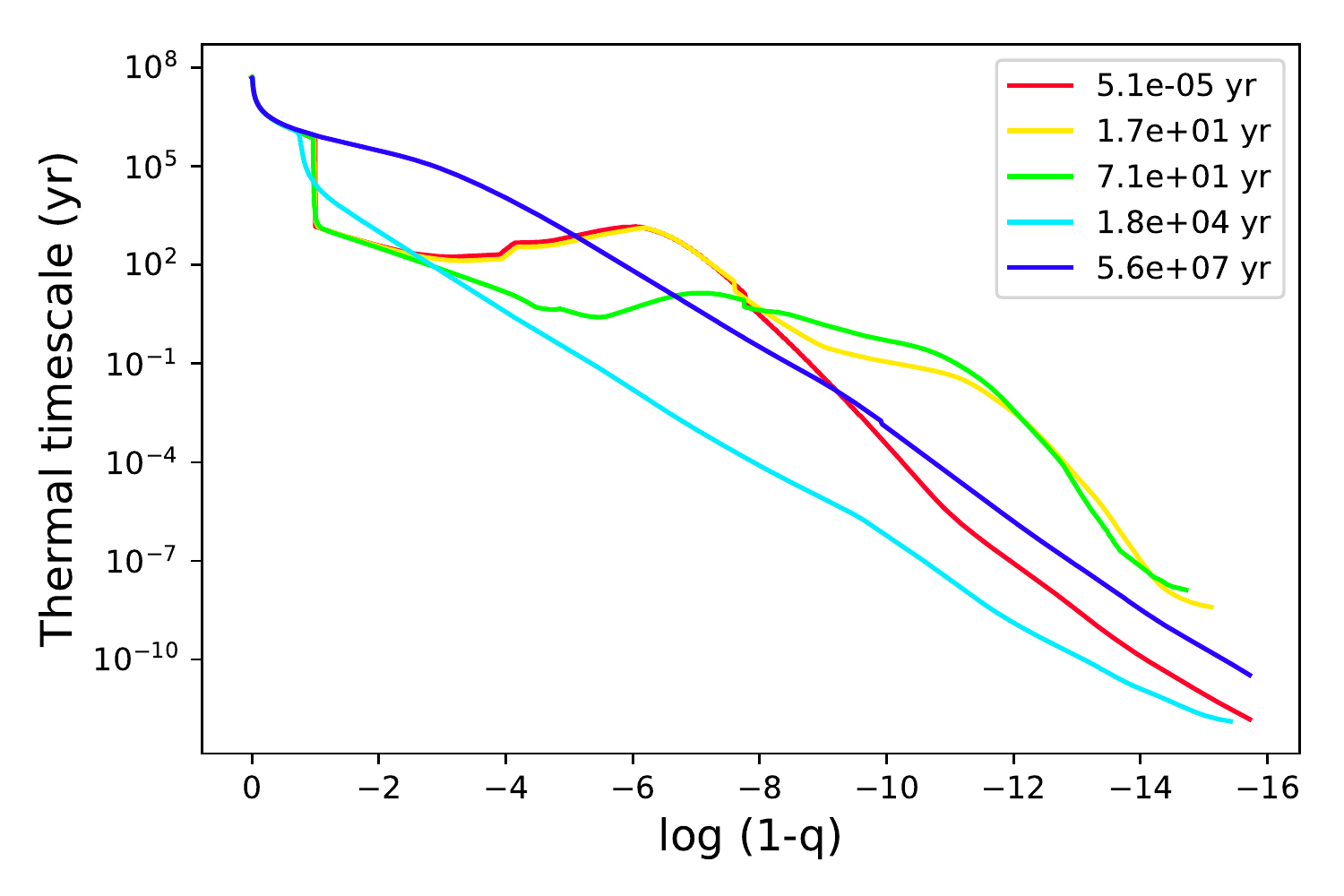}}\subfigure{\includegraphics
    [width=0.5\textwidth]{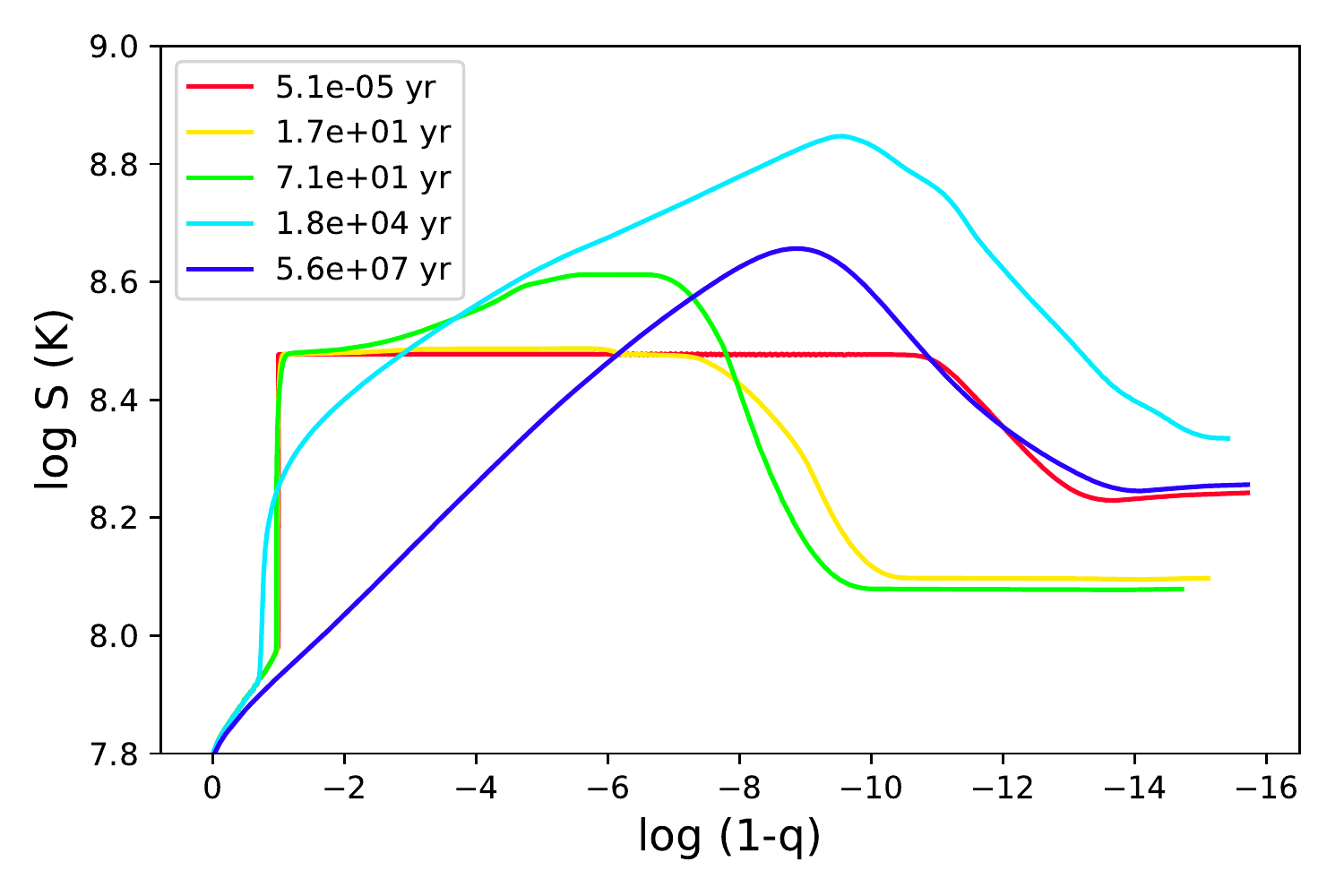}}    
    \caption{Temperature, opacity, thermal timescale, and entropy as a function of depth for selected ages of the 0.6 $M_{\odot}$, low entropy, 10\% envelope fraction model. The mass coordinate of the $x$-axis is $\log(1-q)=\log(1-M_r/M)$, such that the star's surface lies to the right. Colored lines indicate different ages since explosion. From early to late times: at the beginning, the outer layers cool while the interior stays at the same temperature.  Around 71 yr, heat from the interior begins diffusing out, but has not yet reached the surface.  The next profile shows the situation near peak luminosity.  After that, the postgenitor cools. By 56 Myr, which the last profile depicts, the entire envelope has had time to cool, and the postgenitor is well on its way down the cooling track.  Note that the core-envelope transition is at log(1-q) = -1.}
\label{fig:T_kappa_timescale_profiles}
\end{figure*}

The most prominent characteristic of all the models that do not become unbound or swell into red giants is that they have a dimming phase, followed by a rapid re-brightening, followed by a protracted cooling phase akin to those of ordinary WDs.  This decline-rise-decline pattern is robust across a wide range of postgenitor masses, envelope fractions, envelope entropies, and compositions. However, its existence may depend on the constant entropy assumption, which creates a temperature profile that rises sharply with density ($T \propto \rho^{2/3}$), which buries heat deep in the envelope.  Since the thermal timescale increases rapidly with depth, the outer layers cool before the inner layers can react, as shown in Figure \ref{fig:T_kappa_timescale_profiles}. Eventually, heat from the interior diffuses and heats the envelope from the inside out. When the heat reaches the surface, the postgenitor is near peak luminosity. After a thermal time near the star's core, the entire stars cools and descends the WD cooling track.

This behavior is expected so long as we believe the constant entropy assumption.  This is not an obviously bad assumption, as one might expect vigorous mixing in the aftermath of a supernova, which would flatten out the entropy gradient.  Nevertheless, the large unknowns in the explosion mechanism prevents us from proving the assumption is accurate, so the initial decline and subsequent rebrightening should be embraced cautiously.

To test the sensitivity of our models to the constant entropy assumption, we created a model with a 0.6 $M_{\odot}$ WD, 10\% envelope fraction, and the following entropy profile:
\begin{align}
    \frac{d\ln(s)}{dr} = \frac{0.02}{H},
\end{align}
where S is the entropy per unit mass, H is the local scale height, and s=2\e{8} erg g\textsuperscript{-1} K\textsuperscript{-1} at the bottom of the envelope.  This model has a qualitatively similar evolution to the constant entropy model. There is a similar decline-rise-decline pattern, with a uniform composition profile during the decline phase, a photosphere dominated by heavy elements at peak luminosity, and gravitational settling taking over during the cooling phase.  We conclude that our results are not sensitive to the exact shape of the entropy profile.

\subsection{Postgenitor luminosity and evolutionary time scales}

The evolution of the postgenitor is largely controlled by the radiative diffusion of heat out of the deep interior, which is determined by the opacity structure of the star. A unique feature of our postgenitor models is their relatively large abundances of iron group elements in their outer layers. The high opacities created by these elements, coupled with the unusual initial conditions (constant envelope entropy, high iron content) of our models, creates the characteristic dimming and brightening evolution described above.

The initial dimming phase is easy to understand. Heat is transported outward by convection on a thermal timescale $t_{\rm therm}$, causing the outer layers of the star to cool. As the cooling front moves inward, $t_{\rm therm}$ at the base of the cool envelope increases (see red curve in Figure \ref{fig:T_kappa_timescale_profiles}), and so the emerging luminosity decreases with time. This behavior continues until the cooling front reaches a point in the star where $t_{\rm therm}$ has a local maximum. This maximum can be easily seen at $\log(1-q) \sim -6$ in the thermal timescale subplot of Figure \ref{fig:T_kappa_timescale_profiles}.  The heat influx into this layer from underlying layers increases the entropy, setting up a positive entropy gradient and hence radiative energy transport. In our models, $t_{\rm therm}$ initially has a local maximum fairly deep in the star near the iron opacity peak at temperatures of $T \sim 10^6 \, {\rm K}$, which is especially important due to the high iron abundance of our models. 

Eventually, heat diffusing into this layer from below raises its temperature substantially, thereby decreasing its opacity which scales approximately as $\kappa \propto T^{-3.5}$, allowing more heat to diffuse from below. The layer heats more, further decreasing its opacity, causing a runaway process so that a heating wave runs through the envelope toward the surface of the star. The photospheric temperature and luminosity increase suddenly, as shown in Figures \ref{fig:log_Teff_vs_time} and \ref{fig:log_L_vs_time}. The luminosity remains large for roughly one thermal time at the base of the high entropy envelope, after which the star steadily descends the WD cooling track.

The timescale of the WD rebrightening can be estimated via the thermal time in the layers below the iron opacity peak. This timescale is
\begin{align}
t_{\rm therm} &\sim \frac{H^2}{K} \nonumber \\
&\sim \frac{3 H^2 \kappa \rho^2 c_P}{16 \sigma_B T^3},
\end{align}
where $K$ is the thermal diffusivity or thermal diffusion coefficient, $H \sim P/(\rho g)$ is the local scale height,  $\kappa$ is the opacity, $c_P$ is specific heat at constant pressure (computed by MESA from the EOS), and the other variables have their usual meaning.  For bound-free and bound-bound opacity created by iron group elements, the opacity is approximately \citep{hansen_2004}
\begin{equation}
\kappa \sim \kappa_0 \rho T^{-3.5} \, ,
\label{eq:kramers_law}
\end{equation}
where $\kappa_0 \sim 4 \times 10^{25} \, Z \, {\rm cm}^5\,{\rm K}^{3.5}\,{\rm g}^{-2}$ and $Z$ is the metallicity. Deep in the star, $H \sim r$, where $r$ is the local radial coordinate, $\rho \sim 3 M/ 4 \pi r^3$, and the temperature can be approximated from the virial relation,
\begin{equation}
T \sim \frac{G M \mu m_p}{k_B r} \,
\label{eq:virial_T}
\end{equation}
where $\mu$ is the mean molecular weight. This virial relation holds because the high entropy envelope of our WD models is non-degenerate and well-approximated by an ideal gas.

The relevant brightening time is set by the minimal $t_{\rm therm}$ in layers below the iron opacity bump. Figure \ref{fig:T_kappa_timescale_profiles} demonstrates the peak in $t_{\rm therm}$ near the iron opacity bump, which prevents heat from these layers from diffusing outward. However, at larger depths the opacity is lower and the $t_{\rm therm}$ is shorter, such that heat diffusing from deeper in the star warms the gas in the iron opacity bump. Combining the above relations, this happens on the thermal timescale:
\begin{align}
\label{ttherm}
t_{\rm therm} & \sim\frac{81}{256 \pi^3} \frac{k_B^8 \kappa_0}{a c G^7 \mu^8 m_p^8} M^{-4} T^{1/2}   \nonumber \\
& \sim 3 \! \times \! 10^3 \, {\rm yr} \, \bigg(\frac{\mu}{1.75} \bigg)^{\! -8} \bigg( \frac{M}{0.5 \, M_\odot}\bigg)^{\! -4} \bigg(\frac{T}{10^7 {\rm K}}\bigg)^{\! 1/2} \, .
\end{align}
Equation \ref{ttherm} provides a crude estimate of the diffusion timescale that corresponds to the age at which the luminosity increases in Figures \ref{fig:HR_diagrams} and \ref{fig:log_L_vs_time}. The iron opacity peak is around $T_{\rm Fe} \sim 10^6 \, {\rm K}$ at the densities present in our mid-envelope, so we should evaluate equation \ref{tdif} at somewhat warmer temperatures of $\sim 10^7 \, {\rm K}$. The brightening age is smaller in higher mass postgenitors, largely because the internal temperatures are larger such that the iron opacity peak lies closer to the surface where the density is lower and the diffusion time is smaller. In equation \ref{ttherm}, the appropriate mass is the core mass, $(1-f_{\rm env})M$, such that smaller envelope fractions have faster evolution timescales.

The timescale of the peak luminosity in Figure \ref{fig:log_L_vs_time} is given by the photon diffusion time near the base of the high entropy envelope,
\begin{equation}
t_{\rm dif} \sim \frac{r^2 \rho \kappa}{c} \,
\label{tdif}
\end{equation}
Replacing the density, temperature, and opacity as done above, this equates to
\begin{equation}
t_{\rm dif} \sim \frac{9}{16\pi^2}\frac{k_B^{7/2} \kappa_0}{(G \mu m_p)^{7/2}c} M^{-3/2} r^{-1/2} \, .
\label{tdif2}
\end{equation}

Because the luminosity of the postgenitor is powered by gravitational energy release as it contracts into a WD, we must evaluate equation \ref{tdif} where the gravitational energy release is largest, i.e., where $P_{\rm gas} \sim P_{\rm deg}$. At this location, we find the usual WD scaling relation 
\begin{equation}
r \sim \frac{h^2}{G (\mu_e m_p)^{5/3} m_e} M^{-1/3} \, .
\label{eq:wd_mass_radius}
\end{equation}
Combining equations \ref{tdif2} and \ref{eq:wd_mass_radius}, we have
\begin{align}
t_{\rm peak} &\sim t_{\rm dif}\\
& \sim \frac{9}{16\pi^2}\frac{k_B^{7/2} (\mu_e m_p)^{5/6} m_e^{1/2} \kappa_0}{G^3 \mu^{7/2} m_p^{7/2} h c} M^{-4/3} \\
&\sim 7000 \, {\rm yr} \, \bigg(\frac{M}{0.5 M_\odot}\bigg)^{-4/3} \, .
\end{align}
The corresponding peak luminosity is simply the postgenitor's gravitational binding energy divided by the diffusion timescale,
\begin{align}
L_{\rm peak} & \sim \frac{E_{\rm bind}}{t_{\rm dif}} \sim \frac{G M^2}{R t_{\rm dif}} \nonumber \\
& \sim 8000 \, L_\odot \, \bigg( \frac{M}{0.5 \, M_\odot} \bigg)^{\! 11/3} \, .
\end{align}
Although crude, these estimates approximately predict the timescale and luminosity of the peaks in Figures \ref{fig:log_Teff_vs_time} and \ref{fig:log_L_vs_time}, and more importantly, they largely explain the steep scalings with postgenitor mass, which are due to the larger binding energies and lower opacities in more massive postgenitors.

\subsection{Radiative levitation}
\label{radiative}

During the hot and bright phase of our models, radiative levitation becomes strong enough to drive iron and nickel towards the surface, making them the most abundant elements at the surface. These elements are preferentially levitated because they have the most abundant absorption lines, and thus the highest momentum transfer cross sections. Strong radiative levitation of other rare elements (e.g., strontium or tellurium) is also probable, but we do not include these elements in our grid because the Opacity Project does not provide opacities for them.  The transition between a heavy element photosphere and a light element photosphere is abrupt, as can be seen in Figure \ref{fig:surface_abundances}.  For a 0.6 $M_{\odot}$ WD, the transition occurs around $T_{\rm eff}$ = 100,000 K.  This critical temperature drops to 50,000 K for a 0.3 $M_{\odot}$ WD, and climbs to 250,000 K for a 1 $M_{\odot}$ WD.  (The 0.15 $M_{\odot}$ WDs in our grid do not become hot enough for levitation.)  The transition temperature has a simple physical explanation: it is the point at which $g_{\rm rad} = g$ at the photosphere for a given element.

\begin{figure}
  \includegraphics [width= 0.5\textwidth]{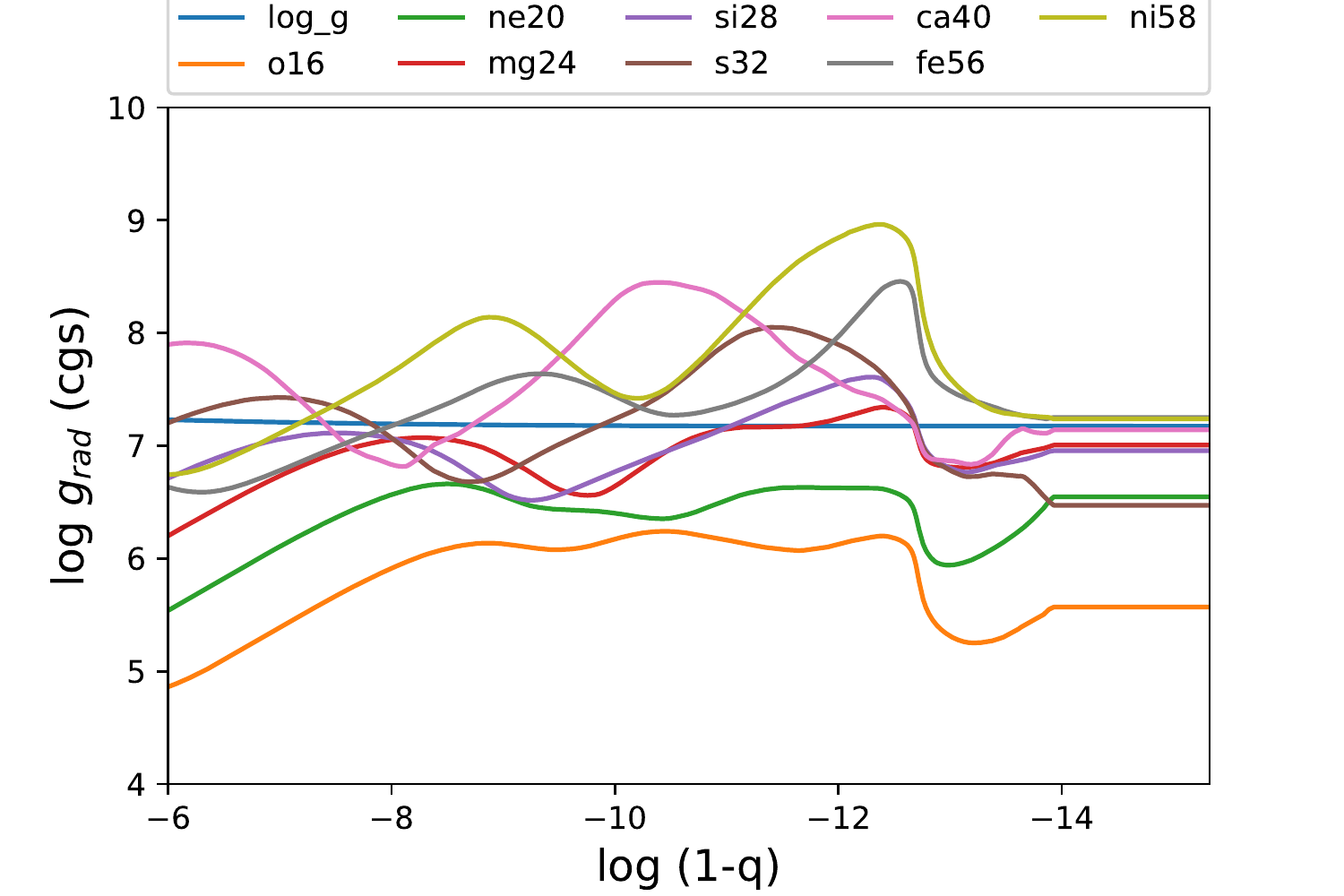}
  \caption{The onset of radiative levitation.  This plot shows $g_{\rm rad}$ as a function of depth for all elements in the 0.6 $M_{\odot}$, 10\% envelope fraction, low entropy model.  This snapshot was taken at 218 yr, when $T_{\rm eff}$ = 150,000 K.}
\label{fig:radiative_levitation}
\end{figure}

\begin{figure}
  \includegraphics [width= 0.5\textwidth]{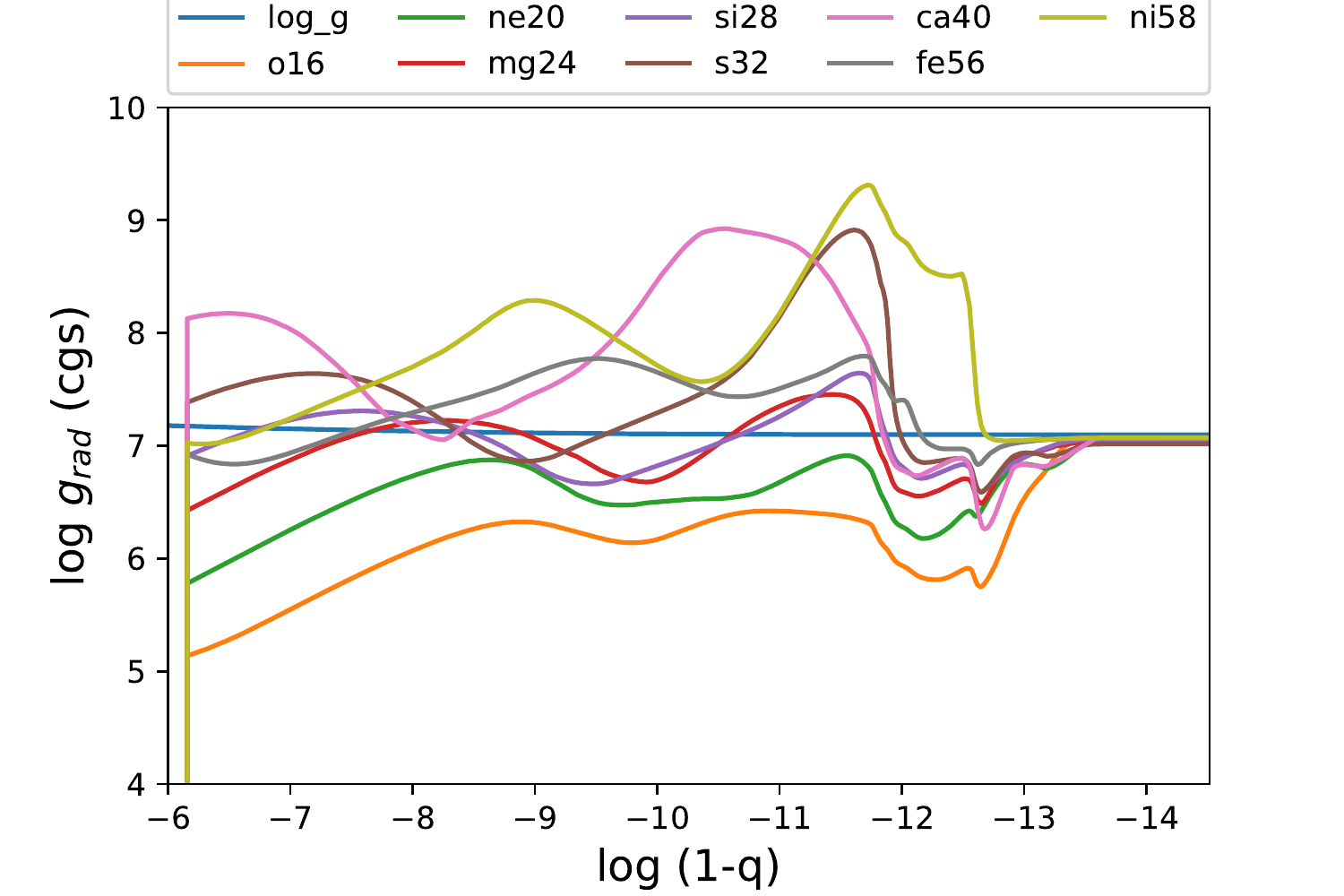}
  \caption{Same as Figure \ref{fig:radiative_levitation}, but at peak brightness.  Notice that radiative acceleration for all elements is close to $g$ at the photosphere.  This snapshot was taken at 2600 yr, when $T_{\rm eff}$ = 200,000 K.}
\label{fig:radiative_levitation_peak}
\end{figure}

\begin{figure}
  \includegraphics [width= 0.5\textwidth]{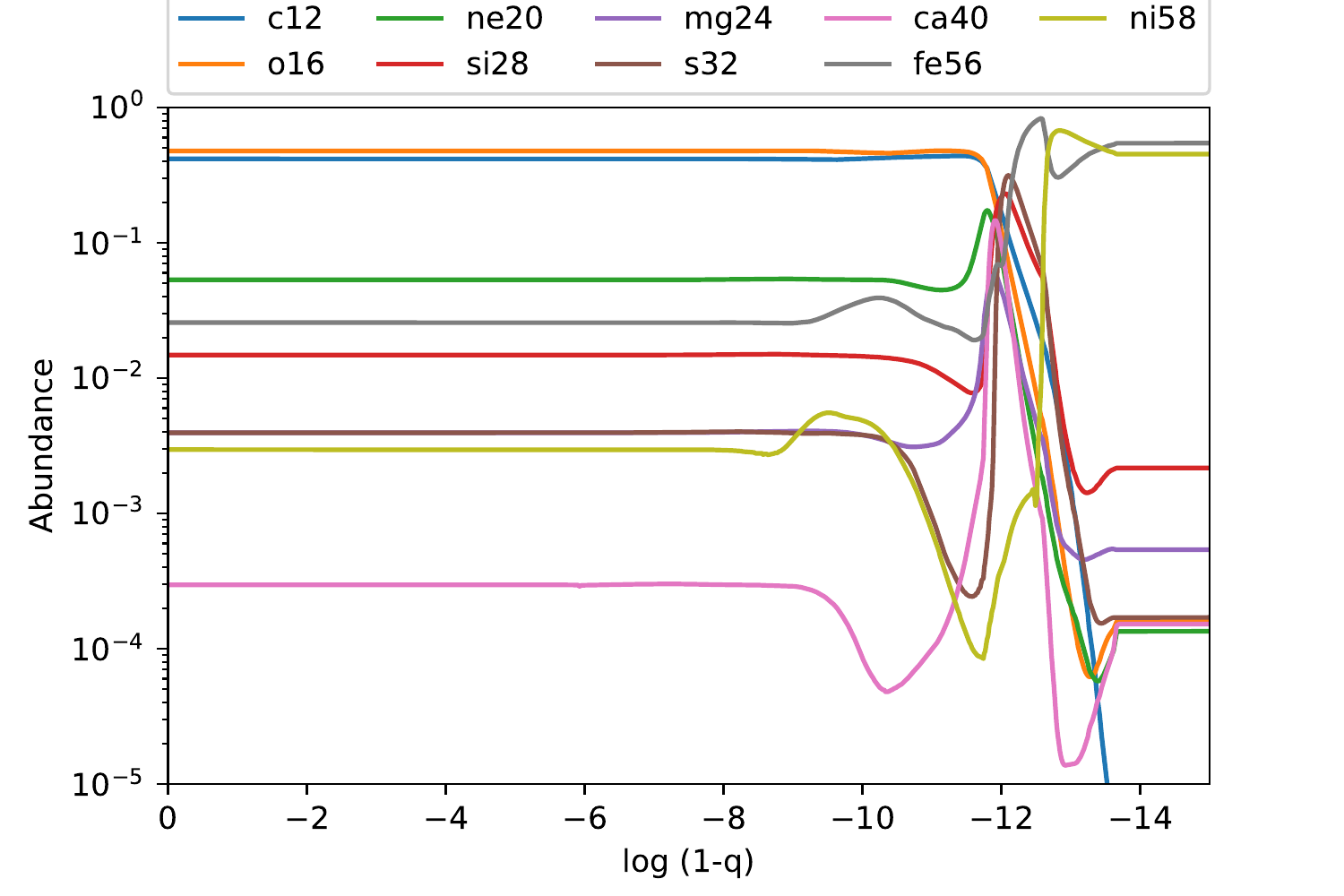}
  \caption{Abundance profile at 2640 yr, corresponding to the same model and timestep as Figure \ref{fig:radiative_levitation_peak}. Note the high surface abundance of iron and nickel due to strong radiative levitation.}
\label{fig:abundance_profile}
\end{figure}

\begin{figure}
  \includegraphics [width= 0.5\textwidth]{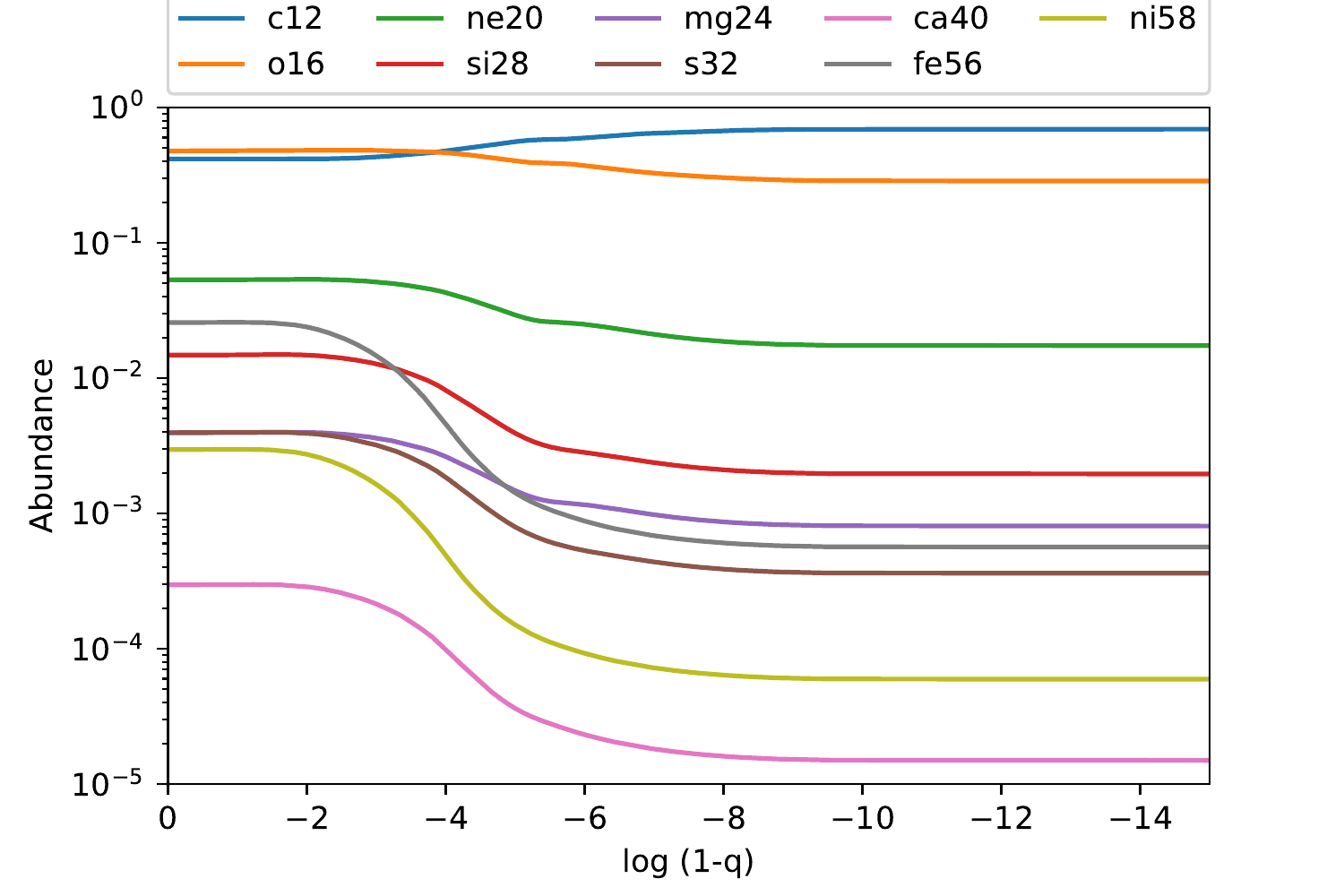}
  \caption{Abundance profile for the model from Figure \ref{fig:abundance_profile},  at very late times (56 Myr) when the WD has cooled.}
\label{fig:abundance_profile_late}
\end{figure}

Figure \ref{fig:radiative_levitation} shows radiative acceleration as a function of position in the star for the hottest model in Figure \ref{fig:surface_abundances}. It can be seen that at this point in time, iron and nickel have $g_{\rm rad} > g$ while the other elements do not.  Not surprisingly, these two are by far the most abundant elements in the photosphere.  As time passes, iron and nickel increase in abundance until $g_{\rm rad} = g$, which occurs $\sim$ 20 yr after the profile shown.  Around peak brightness (2600 yr), $g_{\rm rad}$ for the other elements also approach $g$ (Figure \ref{fig:radiative_levitation_peak}), and their surface abundances reach an equilibrium (Figure \ref{fig:abundance_profile}).  Then the WD cools, and all radiative accelerations drop below gravitational acceleration at $\sim$18,000 yr.  Nickel is still overabundant in the photosphere after this, but the photosphere becomes carbon dominated rather than nickel/iron dominated.  Gravitational settling takes over, and all sign of heavy-element over-abundance is gone by 500,000 yr (Figure \ref{fig:abundance_profile_late}).

One important question is the believability of the surface abundances.  Is it really realistic that the photosphere would be dominated by iron and nickel?  Here we would urge caution.  First, thermohaline diffusion is expected to smooth out composition gradients, but we do not model it. This is discussed more extensively in Subsection \ref{subsec:thermohaline}.  Second, even if thermohaline diffusion were negligible, our method of calculating radiative acceleration assumes the diffusive approximation.  This, in turn, is equivalent to the assumption that the mean free path for a photon is much less than the local atmospheric scale height at all wavelengths.  At the photosphere this assumption is badly violated, and the emergent flux is far from a blackbody.  This is especially the case because radiative levitation is driven by the billions of lines in iron-group elements.  These lines could saturate, leaving no flux at those wavelengths to levitate atoms.  It is also possible that clumps of metal over-densities will form, with most of the flux escaping through gaps between the clumps, and the clumps themselves shielding most atoms inside from radiative levitation.

An additional complication is that when the luminosity approaches or exceeds the Eddington limit--which occurs for most of our higher mass models--the atmosphere develops an inhomogenous porous structure and the effective opacity is greatly reduced.  This effect has been suggested for many super Eddington scenarios, including nova outbursts \citep{kato_2005} and supermassive stars \citep{shaviv_2000}.  Three-dimensional hydrodynamic simulations of radiation-dominated massive star envelopes \citep{jiang_2015,jiang_2017} reveal a complex set of phenomena in super-Eddington atmospheres including shocks, porous atmospheres, and oscillations. MESA uses 1D models and cannot accurately model this porosity, which, by reducing the effective opacity, may also reduce the radiative acceleration. Nor do our MESA models take winds into account.  It is known that mass loss strongly hampers the effects of diffusion and radiative levitation \citep{unglaub_1998,matrozis_2016} by removing levitated elements and pushing the convective zone deeper into the star.

The observational evidence for radiative acceleration indicates that extreme over-abundances of heavy elements are possible, but a photosphere dominated by heavy elements is not.  \cite{werner_2017} took UV spectra of two extremely hot DO WDs ($T_{\rm eff} = 115,000 K$ and $T_{\rm eff} = 125,000\,{\rm K}$) with moderate surface gravity ($\log g = 7 \pm 0.5$), finding that carbon, oxygen, and nitrogen had subsolar abundances while Ne, Si, P, S, Ar, Fe, and Ni have near-solar abundances.  They interpret these abundances as the result of mass loss hampering radiative acceleration.  \cite{werner_2018} searched for metals in hot WDs ($T_{\rm eff} = 65,000 - 120,000\,{\rm K}$) and found light metals with subsolar abundances and iron-group elements with 1-100x solar abundances, which they interpret as the result of gravitational settling and radiative levitation.  \cite{hoyer_2018} searched for trans-iron elements in hot DO WDs and found very high abundances, indicating that radiative levitation is acting.  The most extreme example, PG 0109+111, has a tellurium abundance six orders of magnitude greater than solar.  At a mass abundance of 6.2\e{-3}, it is the most abundant metal in the photosphere.  The fact that hot WDs have been detected with extreme trans-Fe over-abundances but sub-solar intermediate mass element abundances indicates that radiative levitation is not completely overpowered by thermohaline diffusion.

Radiative levitation is certain to be an important effect in the luminous phase of our WDs, and we expect over-abundances of heavy elements.  However, heavy elements are unlikely to become the dominant component of the atmosphere.

\subsection{Candidate Iax postgenitor stars}

Several peculiar WDs have been recently discovered that could be Iax postgenitors. \cite{kepler_2016} discovered a WD with an oxygen-dominated photosphere (SDSS J124043.01+671034.68) with no trace of carbon. However, since carbon burning is required to produce a Iax deflagration, but carbon burning is incomplete in such failed explosions, Iax postgenitors are likely to have substantial carbon abundances. Radiative levitation and gravitational settling are unlikely to eliminate carbon from the photosphere, so we find it unlikely that J1240+6710 is a Iax postgenitor.  We speculate that it could be the remnant of an oxygen deflagation arising from an accreting ONe WD that nears the Chandrasekhar mass. This scenario is similar to the CO deflagation model we have considered, but beginning with an ONe WD, and would naturally explain the lack of carbon. Other possibilities include a deflagation in a hybrid CONe WD \citep{bravo_2016} or a CO-ONe WD merger \citep{kashyap_2018}, though it seems likely such events would leave some carbon in the bound remnant. 

\cite{shen_2018} used \textit{Gaia} data to discover three hypervelocity WD stars. These stars are broadly similar to LP40-365 (see below) in temperature and luminosity, though very different in composition--LP 40-365 is rich in oxygen/neon with little or no carbon, whereas the three Shen objects have carbon in their atmospheres.  The Shen objects are possible Iax postgenitors, although the authors suggest they are instead the companions to Ia progenitors. In fact, one of them appears to have originated within a supernova remnant, lending credence to an explosive origin. We note here that Ia companions and Iax postgenitors may look very much alike--they are both expected to begin as hypervelocity objects with high entropy envelopes, inflated radii, and large abundances of iron group elements. Even though the goal of our paper is to model SNe Iax, our models may turn out to be applicable to Ia companions as well.

LP 40-365 is a peculiar hypervelocity WD (galactocentric velocity = 852 km s\textsuperscript{-1})  with peculiar abundances, originally discovered by \cite{vennes_2017}.  The most abundant photospheric elements are oxygen and neon, followed by intermediate mass elements, while  iron and nickel are detected at a number fraction of $\sim 10^{-3}$.  The authors propose that LP 40-365 is the postgenitor of an exploding carbon-oxygen-neon core. Using Gaia data, \cite{raddi_2018} measured the properties summarized in Table \ref{table:lp_properties}.  They confirmed the hypervelocity nature of the object, measured an abnormally large radius of 0.18 $R_{\odot}$, and found that it crossed the Galactic disk $5.3 \pm 0.5$ Myr ago.  This does not prove that the supernova happened 5.3 Myr ago, as Type Ia supernovae can be found at significant offsets from their host galaxies.  However, \cite{raddi_2018} note that LP 40-365 would have been 100 kpc away 140 Myr ago, setting an upper limit on its age.

\begin{table}[ht]
  \centering
  \caption{LP 40-365 properties, from \cite{raddi_2018}}
  \begin{tabular}{c C}
  \hline
  	Property & \text{Value} \\
      \hline
    $T_{\rm eff}$ & 8900 \pm 300 \\
    $\log g$ & 5.5 \pm 0.25 \\
    $R (R_{\odot})$ & 0.18 \pm 0.01 \\
    $M (M_{\odot})$ & 0.37_{-0.17}^{+0.29}\\
    $L (L_{\odot})$ & 0.18 \pm 0.01\\
    Velocity & 852 \pm 10 \text{ km s\textsuperscript{-1}}\\
    Age & > 5 \text{Myr?}\\
      \hline
  \end{tabular}
  \label{table:lp_properties}
\end{table}

\begin{figure}
  \includegraphics [width= 0.5\textwidth]{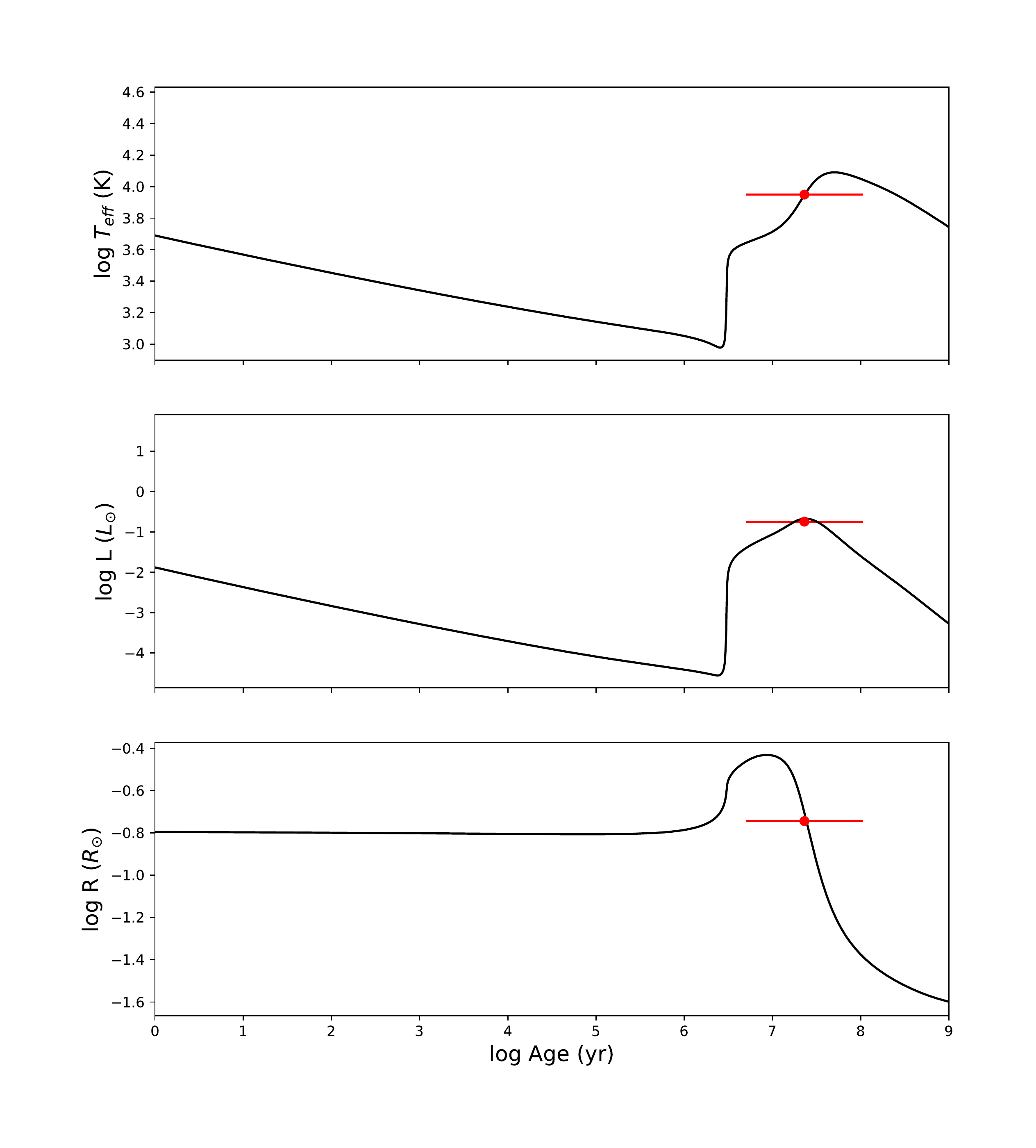}
  \caption{Evolution of the 0.15 $M_{\odot}$ model with 50\% envelope fraction and 3\e{8} erg g\textsuperscript{-1} K\textsuperscript{-1} envelope specific entropy.  The red crosses indicate the observed properties of LP 40-365 at an assumed age of 23 Myr.  The actual age is unknown, but probably between 5 and 100 Myr (indicated by the error bars).  The error bars on the measurements are plotted, but are too small to be seen}
\label{fig:replicating_LP}
\end{figure}

Curious about whether we could explain some of the observed properties with our models, we looked through our grid to find the model that most closely matches LP 40-365.  This turned out to be the 0.15 $M_{\odot}$ model with 50\% envelope fraction and 3\e{8} erg g\textsuperscript{-1} K\textsuperscript{-1} envelope specific entropy, shown in Figure \ref{fig:replicating_LP}.  After a long dimming phase lasting millions of years, this object experiences a broad peak in temperature and luminosity that places it close to LP 40-365 on the HR diagram.

\begin{figure}
  \includegraphics [width= 0.5\textwidth]{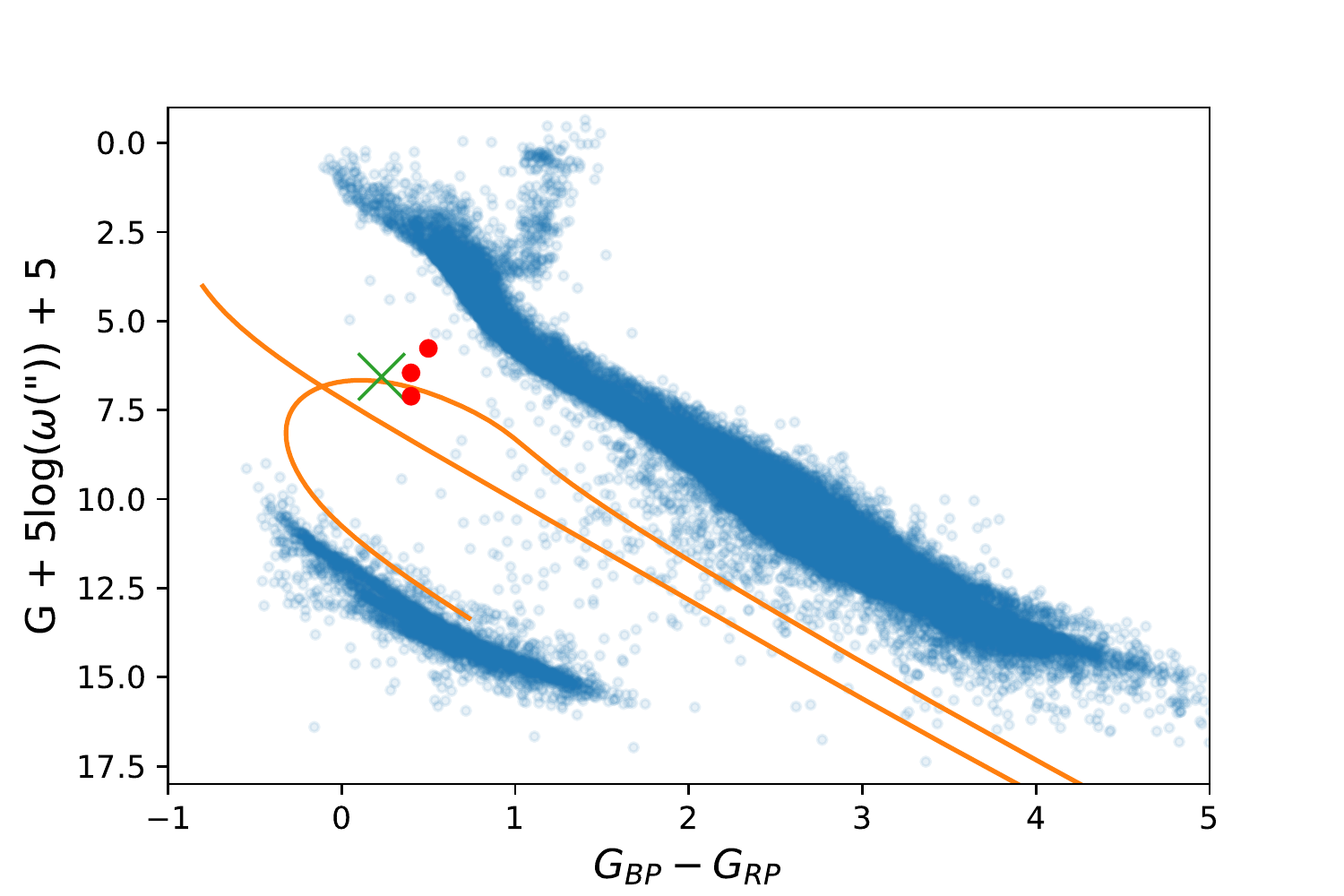}
  \caption{Color-magnitude diagram with Gaia BP-RP colors and G magnitudes.  Green: Iax postgenitor candidate LP 40-365.  Red: the three candidates from \cite{shen_2018}. Orange: our 0.15 $M_{\odot}$ model with 50\% envelope fraction and 3\e{8} erg g\textsuperscript{-1} K\textsuperscript{-1} envelope specific entropy, converted to Gaia quantities by assuming a blackbody spectrum.  Blue: Gaia objects within 100 pc, for reference. Note that our model stays slightly above the main white dwarf cooling track at old age, due to its exceptionally low mass and consequently high radius.}
\label{fig:color_mag_from_Gaia}
\end{figure}

In fact, our model comes strikingly close to matching the observed properties in Table \ref{table:lp_properties}, as can be seen from the color-magnitude diagram in Figure \ref{fig:color_mag_from_Gaia}.  At an age of 23 Myr, our model has a luminosity of 0.21 $L_{\odot}$, a radius of 0.19 $R_{\odot}$, and a temperature of 8977 K. 23 Myr ago, LP 40-365 would have been $\sim$2 kpc from the Galactic disc and $\sim$25 kpc from the Galactic center, reasonable for a SN Iax.  We note that 23 Myr is close to the peak in luminosity and temperature for this model, where evolution is slow, and the WD can linger in this region of the HR diagram for tens of millions of years.  If LP 40-365 actually originated from the disk and is 5 Myr old, a model with slightly higher mass (e.g. 0.2 $M_{\odot}$) would be needed to match its properties, as a more massive WD experiences faster evolution. Despite this close match, we do not claim that our model exactly explains LP 40-365. A key feature of LP 40-365 that disfavors the Iax postgenitor scenario is the low abundance of C in the photosphere, whereas in our models at this temperature, C is the most abundant element. This could indicate that LP 40-365 is better explained as the donor star in a binary with an accreting WD \citep{shen_2018}, or perhaps a partially burnt O-Ne WD \citep{jones_2016}.  Nevertheless, we have shown that it is natural for Iax postgenitor models to match the temperature, luminosity, and radius of LP 40-365 at a reasonable age.

Aside from our CO WD simulations, we also performed an exploratory simulation with an oxygen/neon composition, matching that of LP 40-365.  It encountered numerical problems, but broadly matched the behavior of the C/O white dwarfs in terms of their decline-rise-decline pattern and extreme element over-abundances caused by radiative levitation.

\subsubsection{Number of Detectable Postgenitors}

We can estimate the number of detectable postgenitors assuming each type Iax SN produces a high velocity remnant star. The SNe Iax occurrence rate is $\sim$1/3 that of the SNe Ia occurrence rate \citep{foley_2013}. For a galactic SNe Ia rate of 1 per $\sim$300 years, a SN Iax would occur every $\sim$1000 years. The farthest detected object from \cite{shen_2018} has a distance of $\sim$ 2 kpc. Assuming hypervelocity postgenitors are ejected from the disk at $\sim$1000 km/s, they would travel $2 \, {\rm kpc}$ in 2 Myr, so we expect $\sim$2,000 Iax postgenitors within 2 kpc of the Milky Way's disk.  

Of course, not all of these would be detectable. The number of stars in the disk within the $r= 2 \, {\rm kpc}$ detection volume is $N \approx \pi r^2 h n \approx 6\e{8}$, where h=350 pc is the galactic disk scale height, and n=0.14 pc\textsuperscript{-3} is the local stellar density. Assuming the Milky Way has $2.5 \times 10^{11}$ stars, roughly $0.2 \%$ are within the detection volume. Then we expect $\sim$4 postgenitor stars to be detectable, remarkably similar to the number detected by \cite{shen_2018} and including LP 40-365. While the uncertainty in our estimates are large, the number of observed hypervelocity WDs may be consistent with the number expected from the SN Iax channel. However, one of the \cite{shen_2018} stars appears to originate from a SN remnant.  Adopting $3 \times 10^4$ yr as a SN remnant lifetime, we expect only 1 in $\sim$70 of detectable postgenitors could be traced back to a SN remnant, such that we expect to see only $\sim 0.06$ postgenitors associated with a SN remnant, potentially in tension with the one object traced back to a remnant by \cite{shen_2018}.

\subsection{Thermohaline mixing}
\label{subsec:thermohaline}
One important effect we have not yet considered is thermohaline mixing.  Thermohaline mixing occurs when a radiative region (defined by the Ledoux criterion) exhibits an inverse composition gradient, i.e., it has layers of high molecular weight on top of layers of low molecular weight.  If a blob of high molecular weight material is displaced downwards and no heat exchange occurs, the blob would be less dense than its surroundings and float back up.  However, if substantial heat exchange occurs, the blob cools and becomes denser than its surroundings, thereby continuing to sink.  In Earth's oceans, thermohaline mixing gives rise to ``salt fingers''--so called because sinking blobs create very salty tendrils, sticking deep into less salty subsurface layers.

In stars and WDs, thermohaline mixing has the effect of introducing mixing into radiative regions where mixing would otherwise be negligible.  This mixing is important in scenarios like planetesimal accretion \citep{bauer_2018} and carbon-enhanced metal-poor stars \citep{stancliffe_2008}, where heavier elements accrete on top of a lightweight atmosphere.  In our scenario, radiative levitation tends to push heavy elements upwards, while gravitational settling and thermohaline mixing counteract levitation.

To explore the effect of thermohaline mixing on our WDs, we ran a simulation with thermohaline diffusion enabled.  As discussed in \cite{traxler_2011,zemskova_2014}, thermohaline mixing occurs in regions with $1 < R_0 < K/K_\mu $, with $R_0 = (\nabla_{\rm ad} - \nabla)/\nabla_\mu$, the thermal/composition gradients have their usual definition, and $K_\mu$ is the composition diffusivity. In thermohaline unstable regions, we set the thermohaline diffusion coefficient $D_{\rm therm} \simeq C (K - R_0 K_\mu) / (R_0 - 1)$ according to Equation 4 of \cite{vauclair_2012} (itself derived from \citealt{denissenkov_2010}) with their recommended coefficient of $C = 120$. Enabling thermohaline mixing dramatically changes the chemical composition evolution.  The composition remains nearly constant despite radiative levitation, maintaining the same uniform abundances that we start the simulation with.  This occurs because any increase in the abundances of heavy elements due to levitation increases the molecular weight gradient, which enhances thermohaline mixing and mixes the heavy elements back down.  This also explains why calcium, the least abundant element, can steadily increase in abundance while the others cannot--because calcium contributes negligibly to the molecular weight.

Our thermohaline models run into numerical problems some time after they pass peak luminosity, leading to the diffusion solver failing unless the timescales are very small.  Due to these problems, we leave the full exploration of thermohaline mixing under these conditions to future work, but the implication of these findings is that the surface abundance enhancement of iron group elements will be strongly reduced by thermohaline mixing relative to the predictions of Section \ref{radiative}. 

We can also analytically estimate the equilibrium composition gradient by equating a radiative levitation timescale $t_{\rm rad} = (v_{\rm rad} d\ln \mu/dr)^{-1} $ to a thermohaline mixing timescale $t_{\rm therm} = (D_{\rm therm} d^2 \ln \mu/dr^2)^{-1}$. The length scale $\ell$ on which we expect the composition to vary is then
\begin{equation}
\label{Leq}
    \ell \sim \bigg( \frac{C K H}{v_{\rm rad} (\nabla - \nabla_{\rm ad})} \bigg)^{1/2} \, .
\end{equation}
In our models, equation \eqref{Leq} predicts $\ell \ll H$, so we expect radiative levitation to produce very weak composition gradients when competing with thermohaline mixing, in accordance with the results of our MESA models.  However, we note that rotation and magnetic fields, which are not included, may limit the effects of thermohaline diffusion.  We leave a realistic assessment of these effects to future work.

\section{Conclusion}

We have modeled type Iax supernova postgenitor stars with MESA with a range of initial conditions, accounting for uncertainties in their masses and post-explosion structure.  Not surprisingly, we obtained a wide range of behaviors. Most of our models followed a canonical behavior, starting as hot WDs with abnormally high radii that initially cool and dim. Later, as heat leaks out of the deeper interior, the envelope opacity is reduced, allowing faster radiative diffusion. The stars then become much hotter and brighter on timescales of years to millions of years after the supernova, depending on the star's core and envelope mass.  At peak brightness, all but the lightest WD models have over-abundances of iron group elements in their photospheres due to radiative levitation.  Afterwards, the WDs shrink in radius, cooling and dimming similar to normal WD cooling sequences. Our highest entropy models became unbound, super-Eddington, or inflate into red giants, indicating that some Iax postgenitors could appear as luminous cool stars rather then hot blue stars.

Although the prospect for observing these postgenitors in the early aftermath of a SN Iax is remote, it is not unlikely that a known WD inside the Milky Way is such a postgenitor.  In fact, we already have four candidates, including LP 40-365, which our lowest mass models naturally mimic in luminosity and temperature at a plausible age. Future models for such stars can be improved with a better implementation of thermohaline mixing and mass loss, and realistic estimates for the post-explosion structure. As these models improve, we encourage further deeper observational searches for peculiar WDs and subdwarf remnant stars of various flavors of thermonuclear supernovae.

\section{Acknowledgments}

We thank Ken Shen, Evan Bauer, and Lars Bildsten for useful discussions, and we thank the referee for a very constructive and thorough report. This work was supported by the Heising-Simons Foundation through Grant \#2017-274. Support for this work was provided by NASA through Hubble Fellowship
grant \#HST-HF2-51382.001-A awarded by the Space Telescope Science Institute, which is operated by the Association of Universities for Research in Astronomy, Inc., for NASA, under contract NAS5-26555. This research made use of NASA's Astrophysics Data System.

R.J.F. is supported in part by NSF grant AST-1518052, the Gordon \& Betty Moore Foundation, and by a fellowship from the David and Lucile Packard Foundation.

\textit{Software:} MESA, py\_mesa\_reader, numpy, scipy, matplotlib

\section{Appendix}

The inlist used to run our simulations is pasted below.  This inlist takes model.mod, which must be a model of a WD that has been relaxed to the proper composition and entropy. Source code for our custom version of MESA is available at \url{http://www.astro.caltech.edu/~mz/custom_mesa_10000.tar.gz}.

\begin{verbatim}
&star_job
load_saved_model = .true.
saved_model_name = 'model.mod'
change_initial_net = .true.      
new_net_name = 'sn_Ia.net'
kappa_file_prefix = 'OP_gs98'

/ ! end of star_job namelist

&controls
use_Type2_opacities = .true.
Zbase = 0.05

diffusion_use_isolve = .true.
set_min_D_mix=.true.
min_D_mix=1.0
smooth_convective_bdy=.false.

op_mono_data_path = '...'
op_mono_data_cache_filename = '...'

do_element_diffusion = .true.
diffusion_use_cgs_solver = .true.

diffusion_num_classes = 9 
diffusion_class_representative(1) = 'c12'
diffusion_class_representative(2) = 'o16'
diffusion_class_representative(3) = 'ne20'
diffusion_class_representative(4) = 'mg24'
diffusion_class_representative(5) = 'si28'
diffusion_class_representative(6) = 's32'
diffusion_class_representative(7) = 'ca40'
diffusion_class_representative(8) = 'fe56'
diffusion_class_representative(9) = 'ni58'

diffusion_class_A_max(1) = 12
diffusion_class_A_max(2) = 16
diffusion_class_A_max(3) = 20
diffusion_class_A_max(4) = 24
diffusion_class_A_max(5) = 28
diffusion_class_A_max(6) = 32
diffusion_class_A_max(7) = 40
diffusion_class_A_max(8) = 56
diffusion_class_A_max(9) = 58
         
diffusion_v_max = 1d2
diffusion_max_T_for_radaccel = 1d7
diffusion_calculates_ionization = .true.
diffusion_screening_for_radaccel = .true.

diffusion_min_Z_for_radaccel = 1 
diffusion_max_Z_for_radaccel = 28 

max_abar_for_burning = -1

diffusion_dt_limit = 3d5
diffusion_min_dq_at_surface = 1d-15
min_dq = 1d-16
/ ! end of controls
\end{verbatim}

\bibliographystyle{apj} \bibliography{main}

\end{document}